\documentclass[11pt,reqno]{amsproc}
\allowdisplaybreaks
\numberwithin{equation}{section}

\usepackage{cite}
\usepackage{color}
\usepackage{fullpage}
\usepackage{multirow}
\usepackage{graphicx}
\usepackage{textcomp}
\usepackage{subfigure}
\usepackage{enumerate}
\usepackage[debug=false, colorlinks=true, pdfstartview=FitV,
linkcolor=blue, citecolor=blue, urlcolor=blue]{hyperref}
\usepackage[semicolon,square,authoryear]{natbib}

\usepackage[usenames,dvipsnames]{xcolor}
\usepackage[most]{tcolorbox}

\usepackage[doipre={DOI:~}]{uri}

\newtheorem{remark}{Remark}[section]

\usepackage{morefloats}

\newlength{\drop}
\definecolor{amethyst}{rgb}{0.6, 0.4, 0.8}
\definecolor{burgundy}{rgb}{0.5, 0.0, 0.13}

\title{\textbf{Effect of viscous shearing stresses on optimal material designs for flow of fluids through porous media}}

\author{\textbf{T.~Phatak} and \textbf{K.~B.~Nakshatrala} \\
  {\small Department of Civil and Environmental
    Engineering, University of Houston, Texas. \\
    \textbf{Correspondence to:}~knakshatrala@uh.edu}}


\begin{document}

\begin{titlepage}
  \drop=0.1\textheight
  \centering
  \vspace*{\baselineskip}
  \rule{\textwidth}{1.6pt}\vspace*{-\baselineskip}\vspace*{2pt}
  \rule{\textwidth}{0.4pt}\\[\baselineskip]
       {\Large \textbf{\color{burgundy}
       Effect of viscous shearing stresses on optimal material designs \\[0.3\baselineskip]
       for flow of fluids through porous media}}\\[0.3\baselineskip]
       \rule{\textwidth}{0.4pt}\vspace*{-\baselineskip}\vspace{3.2pt}
       \rule{\textwidth}{1.6pt}\\[\baselineskip]
       \scshape
       An e-print of the paper is available on arXiv. \par
       \vspace*{1\baselineskip}
       Authored by \\[\baselineskip]

  {\Large T.~Phatak\par}
  {\itshape Graduate student, University of Houston, Texas 77204.}\\[0.75\baselineskip]

  {\Large K.~B.~Nakshatrala\par}
  {\itshape Department of Civil \& Environmental Engineering \\
  University of Houston, Houston, Texas 77204. \\
  \textbf{phone:} +1-713-743-4418, \textbf{e-mail:} knakshatrala@uh.edu \\
  \textbf{website:} http://www.cive.uh.edu/faculty/nakshatrala}\\[2\baselineskip]
\begin{figure}[h]
	\subfigure[Darcy-Brinkman model]{\includegraphics[scale=0.36]{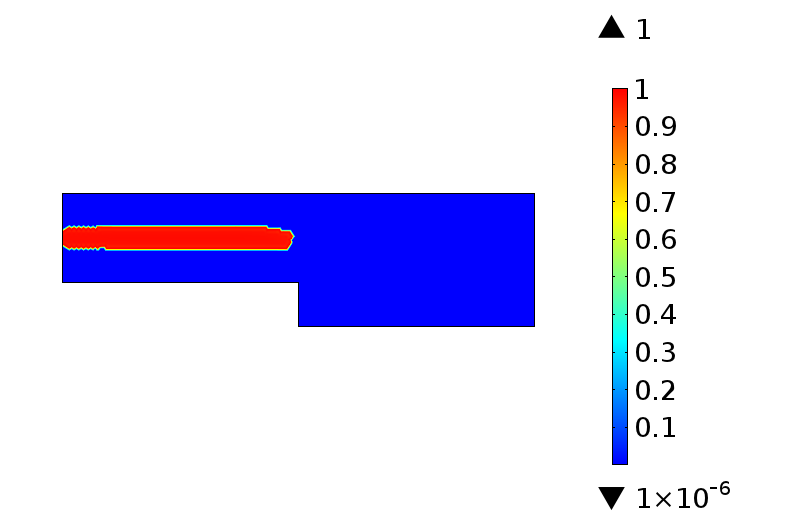}}
	\hspace{0.5cm}
	\subfigure[Darcy model]{\includegraphics[scale=0.36]{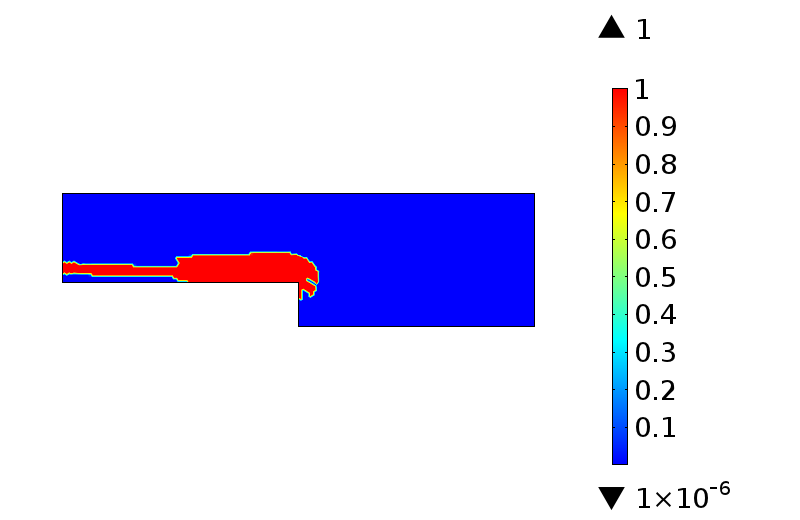}}

	\emph{Using the backward-facing step problem---a benchmark problem in computational
	fluid dynamics---we have shown that the optimal material layouts under the
	Darcy-Brinkman model (left) and the Darcy model (right) differ qualitatively. To drive
	the topology optimization for this pressure-driven problem, we have used the maximization of dissipation
	rate---a physical quantity---as the objective function with a volume
	constraint bound on the high permeability material.}
\end{figure}

  \vfill
  {\scshape 2021} \\
  {\small Computational \& Applied Mechanics Laboratory} \par
\end{titlepage}

\begin{abstract}
Topology optimization offers optimal material layouts, enabling automation in the design of devices. Given the recent advances in computer technology and additive manufacturing, topology optimization is increasingly being used to design complex porous structures, for example, microfluidic devices. For the flow of fluids in such miniature-sized porous structures, viscous shearing stress will be significant. But the Darcy model---the most popular mathematical model describing the flow of a single-phase incompressible fluid in rigid porous media---neglects the internal friction arising from viscous shearing stress. We will therefore develop a material design framework under the topology optimization based on the Darcy-Brinkman model---a mathematical model for the flow of fluids through porous media that accounts for internal friction besides the drag considered in the Darcy model. The proposed framework uses the total rate of mechanical dissipation---a physical quantity---for the objective function. To understand the effect of viscous shearing stress on the design, we will compare the optimal material layouts provided by the Darcy and Darcy-Brinkman models under topology optimization. In particular, we show, using analytical solutions corroborated by numerical simulations, that the optimal material layouts are identical for the class of problems exhibiting axisymmetry, for which viscous shearing stress vanishes. These analytical solutions will be valuable to check the veracity of numerical simulators. For those problems with dominant viscous shearing stresses (e.g., flow in a backward-facing step domain), we show, using numerical solutions, that the optimal material layouts under the Darcy and Darcy-Brinkman models are very different. Also, the associated solution fields (i.e., velocity and pressure) are qualitatively different under their respective optimal layouts for these two models.
\end{abstract}

\maketitle

\vspace{-0.3in}


\section{INTRODUCTION AND MOTIVATION}
\label{Sec:DB_Intro}

Recently, \cite{phatak2020optimal} have studied optimal material designs using topology optimization under the Darcy model and two of the model's nonlinear generalizations: pressure-dependent viscosity (i.e., Barus model) \citep{nakshatrala2011numerical,Barus_AJS_1893_v45_p87} and inertial effects via the modification proposed by \cite{forchheimer1901wasserbewegung}---commonly referred to as the Darcy-Forchheimer model \citep{de2012theory,whitaker1996forchheimer}. The mentioned article has made several advancements in using topology optimization---a mathematically-driven design framework---for flow through porous media applications. Notably, the paper has shown that the principle of minimum power, often used under topology optimization (e.g., see \citep{guest2006topology}), is not valid in general, especially for any nonlinear generalization of the Darcy model. Hence, this principle is not a viable candidate for building a wide-reaching material design framework that is valid for various flow through porous media applications. Alternatively, they provided a topology-optimization-based material design framework using the rate of dissipation; the drag between the fluid and porous solid is taken as the source of dissipation. This framework is shown to be applicable for the two nonlinear generalizations besides the Darcy model. Last but not least, the paper has also outlined the scenarios for which the nonlinear generalizations are preferred over the Darcy model and highlighted the differences in the resulting material designs.

However, Darcy equations and the mentioned generalizations do not invoke no-slip condition at solid surfaces such as impervious boundaries and blunt objects. Mathematically speaking, the governing equations under the three models (i.e., Darcy, Barus and Darcy-Forchheimer), when written in the mixed form, are first-order in terms of the velocity and pressure fields; thus, prescribing the tangential component of the velocity will not render a well-posed boundary value problem. But the no-slip condition along with no penetration condition (i.e., the normal component of the velocity matches that of the solid surface) are well-accepted boundary conditions at an impervious solid surface \citep{batchelor2000introduction}. In short, the Darcy model does not capture the flow conditions accurately near the boundaries and blunt objects.

This issue of compliance with the no-slip condition is crucial for a certain class of microfluidic devices that use porous materials. Microfluidic devices, as the name suggests, realize their functionalities from flow of small volumes of fluid through miniature channels \citep{whitesides2006origins}. These miniature devices have high surface-to-volume ratios and often have solid (or blunt) objects inside devices to the regulate flow patterns within the domain. Notably, the boundaries are often close and their influence on the solution fields is not localized to a small region, but rather affects the whole domain \citep{bruus2008theoretical,tabeling2005introduction}. The first-order models, mentioned above, cannot capture the effect of close-by boundaries and the presence of blunt objects. 

\cite{brinkman1949calculation} realized the inherent drawback of the Darcy model when he wanted to calculate the frictional forces (i.e., the drag force) on a swarm of particles in a fluid flow. Motivated by the work of \cite{stokes1851effect} on free flows, Brinkman modified the Darcy model to include a second-order term in terms of the velocity field. This new model is commonly referred to as the Darcy-Brinkman (or just the Brinkman) model. The new term accounts for fluid's internal friction, thereby captures viscous shearing stresses which are dominant near boundaries and blunt object. Also, the second-order term increases the order of the resulting partial differential equation to two, thus, allowing the prescription of the tangential component of the fluid flow besides the normal component. 

The Darcy-Brinkman model alongside topology optimization provides a way to design microfluidic devices with porous materials that have close-by boundaries and possibly blunt objects within the devices. Although initial works on topology optimization have been towards structural mechanics, there is currently a lot of research activity on applying topology optimization to problems in fluid mechanics and flow of fluids through porous media. To name a few, see \citep{borrvall2003topology, evgrafov2005limits, kreissl2011topology, guest2006topology,wiker2007topology,gersborg2005topology}, the survey article by \cite{andreasen2009topology,deaton2014survey}, and the discussion in \citep{phatak2020optimal}. Topology optimization has been studied for over five decades, in engineering and applied mathematics fields \citep{bendsoe1988generating,bendsoe_sigmund_2013topology,rozvany2014topology}. But this optimization technique has once again come to the forefront because of: (1) New manufacturing techniques (e.g., additive manufacturing) can now fabricate complex material designs, such as the ones provided by topology optimization. (2) The growth in computer power and parallel computing environments make practical, large-scale design problems tractable. (3) Developments on the algorithmic front (e.g., filters, solvers) allow obtaining manufacturable (i.e., 0--1) solutions.

However, if the underlying primal analysis is not accurate, the resulting design will not be optimal. Thus, the selection of the model, which defines the primal analysis, is crucial. The importance of this point cannot be overemphasized, especially when applying topology optimization for applications involving the flow of fluids through porous media, as this field has numerous models. But different models can give rise to different material designs, resulting in entirely different solution fields. Prior studies did not address this paper's primary focus---the effect of viscous shearing stresses on the optimal material distribution.

The current knowledge gap can be posed in form of following questions:
\begin{enumerate}[(Q1)]
\item Would optimal material layouts under the Darcy-Brinkman model differ from those analyzed using the Darcy model? 
\item Do viscous shearing stresses affect the optimal material layout? 
\item Do geometrical features (e.g., abrupt changes) of the domain play a role towards this difference? 
\item Is there any class of problems for which the optimal material layouts under these two models are identical? 
\item If there are differences in the optimal material distributions, how to choose an appropriate (either Darcy-Brinkman or Darcy) model for a given problem? 
\end{enumerate}

Given the importance of miniature-sized devices in the diagnostic/chemical analysis field  \citep{beebe2002physics,weibel2006applications,whitesides2006origins,stone2004engineering} and current potential of topology optimization to provide optimal designs for such devices, there is a \emph{need} to address the above questions. Ignoring these factors can lead to imperfect designs affecting accuracy of porous miniature-sized devices involving fluid flows. The \emph{aim} of this paper is two-fold. First, we will present a material design framework using topology optimization that takes into account the internal friction within the fluid besides the drag between the fluid and porous skeleton. Second, we will gain a better understanding of optimal material distributions involving flow of fluids through porous media by addressing the aforementioned questions. 

Our \emph{approach} is to use the Darcy-Brinkman model, which account for both fluid's internal friction as well as the drag between the fluid and the porous solid, which is the only dissipative mechanism under the Darcy model. Motivated by our recent paper \citep{phatak2020optimal}, we will use the rate of dissipation---a physical quantity with firm thermodynamic underpinning---for the objective function to drive the design problem. We will restrict our study to pressure-driven problems. So, the design problem is to place two given materials, with different permeabilities, within a domain so as to maximize the rate of dissipation with a volume constraint placed on the amount of usage of the high-permeability material. Using the proposed material design framework and a combination of analytical and numerical solutions, we will provide answers to the questions outlined above. 

An outline for the rest of this article is as follows. \S\ref{Sec:S2_Brinkman_GE} documents the balance laws describing the flow of an incompressible fluid through porous media, along with the constitutive equations for Darcy-Brinkman and Darcy models. \S\ref{Sec:S3_Brinkman_Design} presents the mathematical description of the material design problem, built based on topology optimization and the total dissipation rate, that takes into account viscous shearing stress and no-slip condition on solid surfaces. Next, optimal layouts for 2D (\S\ref{Sec:S4_Brinkman_AxiSym_2D}) and 3D (\S\ref{Sec:S5_Brinkman_AxiSym_3D}) axisymmetric problems, for which viscous shearing stress vanishes, are obtained using an analytical approach by assuming a single material interface; this assumption is validated using numerical solutions. After that, optimal material layouts for two representative problems exhibiting significant viscous shearing stresses (pipe-bend problem, \S\ref{Sec:S6_Brinkman_Pipe_bend}, and backward-facing step problem, \S\ref{Sec:S7_Backward_Facing_Step}) will be shown. Finally, conclusions on the nature of optimal solutions under the two models will be drawn (\S\ref{Sec:S8_Brinkman_CR}).

\section{DARCY-BRINKMAN AND DARCY MODELS}
\label{Sec:S2_Brinkman_GE}
Consider a porous domain denoted by $\Omega \subset \mathbb{R}^{nd}$,
where ``$nd$'' denotes the number of spatial dimensions. The domain is
assumed to be bounded by a piecewise smooth boundary $\partial \Omega$.
Mathematically, $\partial \Omega = \overline{\Omega} - \Omega$, where
an overbar denotes the set closure. A spatial point is denoted by
$\mathbf{x} \in \overline{\Omega}$. The gradient and divergence
operators with respect to $\mathbf{x}$ are, respectively, denoted
by $\mathrm{grad}[\cdot]$ and $\mathrm{div}[\cdot]$.

We are concerned with the flow of a single-phase incompressible
fluid through the porous domain. The porous solid is assumed to
be rigid. The density and dynamic coefficient of viscosity of
the fluid are denoted by $\rho$ and $\mu$, respectively. The
specific body force is denoted by $\mathbf{b}(\mathbf{x})$. The
permeability field of the porous solid is denoted by $k(\mathbf{x})
> 0$; determining the actual spatial variation of this field---often
referred to as the material design---is \emph{central} to this paper.

The velocity and the pressure of the fluid in the domain are
denoted by $p(\mathbf{x})$ and $\mathbf{v}(\mathbf{x})$,
respectively. The symmetric part of the velocity gradient
is denoted by $\mathbf{D}$. That is, 
\begin{align}
  \mathbf{D} = \frac{1}{2} \left(\mathrm{grad}[\mathbf{v}]
  + \mathrm{grad}[\mathbf{v}]^{\mathrm{T}}\right)
\end{align}
We denote the unit outward normal vector to the boundary
by $\widehat{\mathbf{n}}(\mathbf{x})$. The boundary is
divided into two complementary parts. The part of the
boundary on which velocity boundary conditions are
prescribed is denoted by $\Gamma^{v}$. $\Gamma^{t}$
denotes that part of the boundary on which traction
boundary conditions are prescribed. For mathematical
well-posedness, we have:
\begin{align}
  \Gamma^{v} \cup \Gamma^{t} = \partial \Omega
  \quad \mathrm{and} \quad
  \Gamma^{v} \cap \Gamma^{t} = \emptyset
\end{align}

The governing equations corresponding to the \emph{Darcy-Brinkman} model take the following form:
\begin{subequations}
  \label{Eqn:Brinkman_GE_Darcy_model}
  \begin{alignat}{2}
    \label{Eqn:Brinkman_BoLM}
    &\frac{\mu}{k(\mathbf{x})} \mathbf{v} + \mathrm{grad}[p]
    - \mathrm{div}[2 \mu \mathbf{D}] = \rho \mathbf{b}(\mathbf{x})
    &&\quad \mathrm{in} \; \Omega \\
    \label{Eqn:Brinkman_BoM}
    &\mathrm{div}[\mathbf{v}] = 0
    && \quad \mathrm{in} \; \Omega \\
    \label{Eqn:Brinkman_vBC}
    &\mathbf{v}(\mathbf{x}) = \mathbf{v}^{\mathrm{p}}(\mathbf{x}) 
    && \quad \mathrm{on} \; \Gamma^{v} \\
    \label{Eqn:Brinkman_traction_BC}
    &\left(-p(\mathbf{x}) \mathbf{I} + 2 \mu \mathbf{D}\right)
    \widehat{\mathbf{n}}(\mathbf{x}) = \mathbf{t}^{\mathrm{p}}(\mathbf{x})
    && \quad \mathrm{on} \; \Gamma^{t} 
  \end{alignat}
\end{subequations}
where $\mathbf{v}^{\mathrm{p}}(\mathbf{x})$ is the prescribed velocity on the boundary, $\mathbf{t}^{\mathrm{p}}(\mathbf{x})$ is the prescribed traction on the boundary, and $\mathbf{I}$ is the second-order identity tensor. The governing equations corresponding to the \emph{Darcy} model take the following form:
\begin{subequations}
  \label{Eqn:Brinkman_GE_Darcy_model}
  \begin{alignat}{2}
    \label{Eqn:Brinkman_Darcy_BoLM}
    &\frac{\mu}{k(\mathbf{x})} \mathbf{v} + \mathrm{grad}[p]
    = \rho \mathbf{b}(\mathbf{x})
    &&\quad \mathrm{in} \; \Omega \\
    \label{Eqn:Brinkman_Darcy_BoM}
    &\mathrm{div}[\mathbf{v}] = 0
    && \quad \mathrm{in} \; \Omega \\
    \label{Eqn:Brinkman_Darcy_vBC}
    &\mathbf{v}(\mathbf{x}) \cdot
    \widehat{\mathbf{n}}(\mathbf{x}) = v_n(\mathbf{x}) 
    && \quad \mathrm{on} \; \Gamma^{v} \\
    \label{Eqn:Brinkman_Darcy_traction_BC}
    &p(\mathbf{x}) = p_0(\mathbf{x})
    && \quad \mathrm{on} \; \Gamma^{t} 
  \end{alignat}
\end{subequations}
where $v_n(\mathbf{x})$ is the prescribed normal component of the velocity on the boundary, and $p_0(\mathbf{x})$ is the prescribed pressure on the boundary.

There are \emph{five} notable differences between the Darcy-Brinkman and Darcy models:
\begin{enumerate}[(i)]
\item \emph{PDE's order}: The order (i.e., the number of
  spatial derivatives) of the partial differential equation
  (PDE) under Darcy-Brinkman equations is two while it is
  one under Darcy equations. See the balance of linear
  momentum given by equations \eqref{Eqn:Brinkman_Darcy_BoLM}
  and \eqref{Eqn:Brinkman_BoLM}. 
\item \emph{Cauchy stress}:
  The Cauchy stress under the two models can be written as follows:
\begin{alignat}{2}
&\mathbf{T} = - p \mathbf{I} &&\quad \mbox{(Darcy model)} \\
&\mathbf{T} = - p \mathbf{I} + 2 \mu \mathbf{D} 
&&\quad \mbox{(Darcy-Brinkman model)}
\end{alignat}
This difference in the Cauchy stress manifests in the balance of linear momentum, and is the reason for the difference in the order of the PDEs. Also, the Cauchy stress under the Darcy model is an isotropic tensor. However, due to the presence of the off-diagonal terms in $2 \mu \mathbf{D}$---the so-called viscous shearing stresses---the Cauchy stress under the Darcy-Brinkman model is, in general, not an isotropic tensor. 
\item \emph{Velocity boundary condition} (cf. equations
  \eqref{Eqn:Brinkman_Darcy_vBC} and \eqref{Eqn:Brinkman_vBC}):
  The Darcy model allows for the prescription of
  only the normal component of the velocity on the
  boundary. On the other hand, one can prescribe
  the entire velocity vector on the boundary
  under the Darcy-Brinkman model.
\item \emph{Traction boundary condition} (cf. equations
  \eqref{Eqn:Brinkman_Darcy_traction_BC} and
  \eqref{Eqn:Brinkman_traction_BC}): One
  can prescribe the pressure in the fluid
  under the Darcy model via prescribing
  a pressure boundary conditions. On
  the other hand, under the Darcy-Brinkman model,
  a prescribed pressure boundary condition is
  equal to the prescribed traction on the boundary;
  the traction (given by $\mathbf{T}\widehat{\mathbf{n}}$)
  consists of the pressure in the fluid as well as shear
  stress. So, the pressure in the fluid under the Darcy-Brinkman
  model need not be equal to the prescribed pressure loading on
  the boundary.  
\item \emph{Rate of dissipation}:
  The Darcy model considers only the drag between the fluid
  and the porous skeleton. Whereas the Darcy-Brinkman model
  also considers the internal friction within the fluid,
arising due to viscous shearing stress. The rate of dissipation
density under the Darcy-Brinkman model is:
  \begin{align}
  \label{Eqn:DB_varphi_Darcy_Brinkman}
    \varphi(\mathbf{x}) = \frac{\mu}{k(\mathbf{x})} \mathbf{v}(\mathbf{x}) \cdot 
    \mathbf{v}(\mathbf{x}) + 2 \mu \mathbf{D} \cdot \mathbf{D} 
  \end{align}
  The rate of dissipation density under the Darcy model is:
  \begin{align}
    \label{Eqn:DB_varphi_Darcy}
    \varphi(\mathbf{x}) = \frac{\mu}{k(\mathbf{x})} \mathbf{v}(\mathbf{x}) \cdot 
    \mathbf{v}(\mathbf{x}) 
  \end{align}
\end{enumerate}

Given the said differences between the two mathematical models, a natural question to ask is: \emph{do the optimal designs under the Darcy-Brinkman and Darcy models differ under the topology optimization}. Specifically, what is the class of problems under which the optimal designs for these models are similar, and what are the scenarios under which the optimal designs are qualitatively different. Also, how do the corresponding solution fields compare under their respective optimal designs for these two models? In the rest of this paper, we will answer these questions using analytical and numerical solutions. 

\begin{remark}
  For the Darcy-Brinkman model, one can find in the literature
  two different ways to handle the applied traction in the form
  of a pressure boundary condition on $\Gamma^{t}$. The first
  way is to enforce the entire traction in terms of the
  prescribed ambient pressure field $p^{\mathrm{p}}(\mathbf{x})$:
  \begin{align}
    \left(-p(\mathbf{x}) \mathbf{I} + 2 \mu \mathbf{D}
    \right) \widehat{\mathbf{n}}
    = \mathbf{t}^{\mathrm{p}}(\mathbf{x}) =
    - p^{\mathrm{p}}(\mathbf{x}) \widehat{\mathbf{n}}(\mathbf{x})
  \end{align}
  The second way is to enforce only the normal component of the traction based on the prescribed pressure and  prescribe the tangential component of the velocity on the same part of the boundary. Mathematically,
  \begin{subequations}
    \begin{align}
      &\widehat{\mathbf{n}}^{\mathrm{T}}(\mathbf{x}) 
      \left(-p(\mathbf{x}) \mathbf{I} + 2 \mu \mathbf{D}
	\right) \widehat{\mathbf{n}}(\mathbf{x}) 
      = - p^{\mathrm{p}}(\mathbf{x}) \\
      &\left(\mathbf{I} - \widehat{\mathbf{n}}(\mathbf{x})
      \otimes \widehat{\mathbf{n}}(\mathbf{x}) 
      \right) \mathbf{v}(\mathbf{x}) = \mathbf{v}_{\|}(\mathbf{x})
    \end{align}
  \end{subequations}
  where $\mathbf{v}_{\|}(\mathbf{x})$ is the component of
  the velocity vector field tangential to the boundary,
  and $\otimes$ denotes the tensor product. If the boundary
  is fixed, we have $\mathbf{v}_{\|}(\mathbf{x}) = \mathbf{0}$.
  The second-order tensor $\mathbf{P}_{\|} := \mathbf{I} -
  \widehat{\mathbf{n}}(\mathbf{x}) \otimes \widehat{\mathbf{n}}(\mathbf{x})$
  is a projection; that is, $\mathbf{P}_{\|} \mathbf{P}_{\|} = \mathbf{I}$.
  The tensor $\mathbf{P}_{\|}$ acts on a vector and projects the vector tangential to the plane defined by the normal vector $\widehat{\mathbf{n}}(\mathbf{x})$. In all our numerical simulations, we have used the second way of enforcing pressure boundary conditions.
\end{remark}

\section{DESIGN PROBLEM USING TOPOLOGY OPTIMIZATION}
\label{Sec:S3_Brinkman_Design}

Topology optimization simulation is driven by defining the objective function and selecting the appropriate extremization---maximization or minimization---of the objective function. Following our prior work \citep{phatak2020optimal}, we will use the rate of mechanical dissipation over the entire domain as the objective function. In this paper, we shall restrict to pressure-driven problems; hence, we will maximize the total rate of dissipation with a volumetric bound constraint on the high-permeability material. However, one can easily extend the studies to velocity-driven problem by minimizing the total rate of dissipation with, again, a volumetric bound constraint, again, on the high-permeability material. The total rate of dissipation is defined as follows:
\begin{align}
  \label{Eqn:DB_total_rate_of_dissipation}
  \Phi = \int_{\Omega} \varphi(\mathbf{x}) \mathrm{d} \Omega
\end{align}
where $\varphi(\mathbf{x})$ is the rate of dissipation density. The expressions for $\varphi(\mathbf{x})$ under the Darcy-Brinkman and Darcy models are provided by equations \eqref{Eqn:DB_varphi_Darcy_Brinkman} and \eqref{Eqn:DB_varphi_Darcy}, respectively. Next, the statement for the material design problem is provided. 

\begin{tcolorbox}
  \begin{center}
    \textbf{Material design problem}
  \end{center}
  Given a domain with prescribed boundary conditions, and two porous materials with different permeabilities, distribute these materials within the domain so as to
  \begin{enumerate}[(a)] 
  \item maximize the total rate of dissipation, 
  \item satisfy the governing equations of the primal analysis (e.g., balance laws and boundary conditions) that accounts for the fluid's internal friction besides the drag between the fluid and porous solid, and 
  \item meet the volumetric bound constraint that limits the area/volume occupied by the high-permeability (i.e., constrained) material. No restriction is placed on the total area occupied by the low-permeability (i.e., unconstrained) material.
  \end{enumerate}
\end{tcolorbox}

We introduce a design field variable $\xi(\mathbf{x})$ which takes either $0$ or $1$ at each spatial point $\mathbf{x}$. The design variable will determine the permeability at a given spatial point; that is, whether the spatial point is occupied by the constrained material (i.e., material 1) or unconstrained material (material 2). Thus, the permeability field takes the following form: 
\begin{align}
  k(\mathbf{x}) = \left\{\begin{array}{ll}
  k_1 & \mathrm{if} \; \xi(\mathbf{x}) = 1 \\
  k_2 & \mathrm{if} \;  \xi(\mathbf{x}) = 0 
  \end{array}\right.
\end{align}
where $k_1$ and $k_2$ denote the permeabilities of materials 1 and 2, respectively.

$\Phi$ is defined through $\varphi$ which depends on the velocity field and the permeability field; the latter is characterized by the design field variable. Also, the velocity field depends on the design variable in an implicit manner. Thus, $\Phi$ is a functional---a function of functions---of $\mathbf{v}(\mathbf{x})$ and $\xi(\mathbf{x})$, and hence will be denoted by $\Phi[\xi(\mathbf{x}),\mathbf{v}(\mathbf{x})]$.  

Mathematically, the material design using the topology optimization can be posed as follows: 
\begin{subequations}
  \label{Eqn:Brinkman_integer_programming_original}
  \begin{alignat}{2}
    &\widehat{\xi}(\mathbf{x}) \leftarrow
    \mathop{\mathrm{argmax}}_{\xi(\mathbf{x})}
    \; \Phi [\xi(\mathbf{x}),
      \mathbf{v}(\mathbf{x})] 
    && \quad \mbox{(objective functional)} \\  \notag
    &\mbox{subject to:} \\
    & \left.
    \begin{array}{ll}
      \frac{\mu}{k(\xi(\mathbf{x}))} \mathbf{v} + \mathrm{grad}[p]
      - \mathrm{div}[2 \mu \mathbf{D}] = \rho \mathbf{b}(\mathbf{x})
      &\quad \mathrm{in} \; \Omega \\  
      \mathrm{div}[\mathbf{v}] = 0 &\quad \mathrm{in} \; \Omega \\  
      \mathbf{v}(\mathbf{x}) = \mathbf{v}^{\mathrm{p}}(\mathbf{x}) 
      & \quad \mathrm{on} \; \Gamma^{v} \\  
      \left(-p(\mathbf{x}) \mathbf{I} + 2 \mu \mathbf{D}\right)
      \widehat{\mathbf{n}}(\mathbf{x}) = \mathbf{t}^{\mathrm{p}}(\mathbf{x})
      & \quad \mathrm{on} \; \Gamma^{t} 
    \end{array} \right\} 
    && \quad \mbox{(state equations)} \\
    &\int_{\Omega} \xi(\mathbf{x}) \, \mathrm{d} \Omega
    \leq \gamma \, \mbox{meas}(\Omega)
    &&\quad \mbox{(volume constraint)} \\
    &\xi(\mathbf{x})\in \left\{0,1\right\} \quad \forall \mathbf{x} \in \Omega
    && \quad \mbox{(design set/space)}
  \end{alignat}
\end{subequations}
where $\xi(\mathbf{x})$ is the design variable, and $\gamma$ is the volume constraint bound for the constrained (sometimes also referred to as the controlled) material. The volume bound constraint, $\gamma$, controls the volume of the constrained material and defines the inequality constraint.

For achieving manufacturable solutions, we use the homogenization method and the SIMP regularization to get (nearly) 0--1 solutions \citep{bendsoe_sigmund_2013topology}, where $1$ indicates the presence of the constrained material while $0$ indicates the unconstrained (or uncontrolled) material. The MMA solver \citep{svanberg1987method} is used to ensure faster convergence of the numerical simulation.

The above design problem is implemented using \cite{COMSOL}---a popular multiphysics finite element-based simulator. In all our numerical simulations, the optimization procedure is terminated if one of the following stopping criteria is met: (i) The design variable is close to a 0--1 solution with no apparent change in the material distribution over several consequent iterations, and the number of iterations is 500. (ii) The relative tolerance for objective function in successive iterations is less than 0.001.

In the next several sections, we will solve the material design problem for various boundary value problems to provide answers to the questions outlined in Introduction (\S\ref{Sec:DB_Intro}).

\section{OPTIMAL MATERIAL DESIGN FOR A 2D AXISYMMETRIC PROBLEM}
\label{Sec:S4_Brinkman_AxiSym_2D}
\begin{figure}[h]
	\includegraphics[scale=0.5]{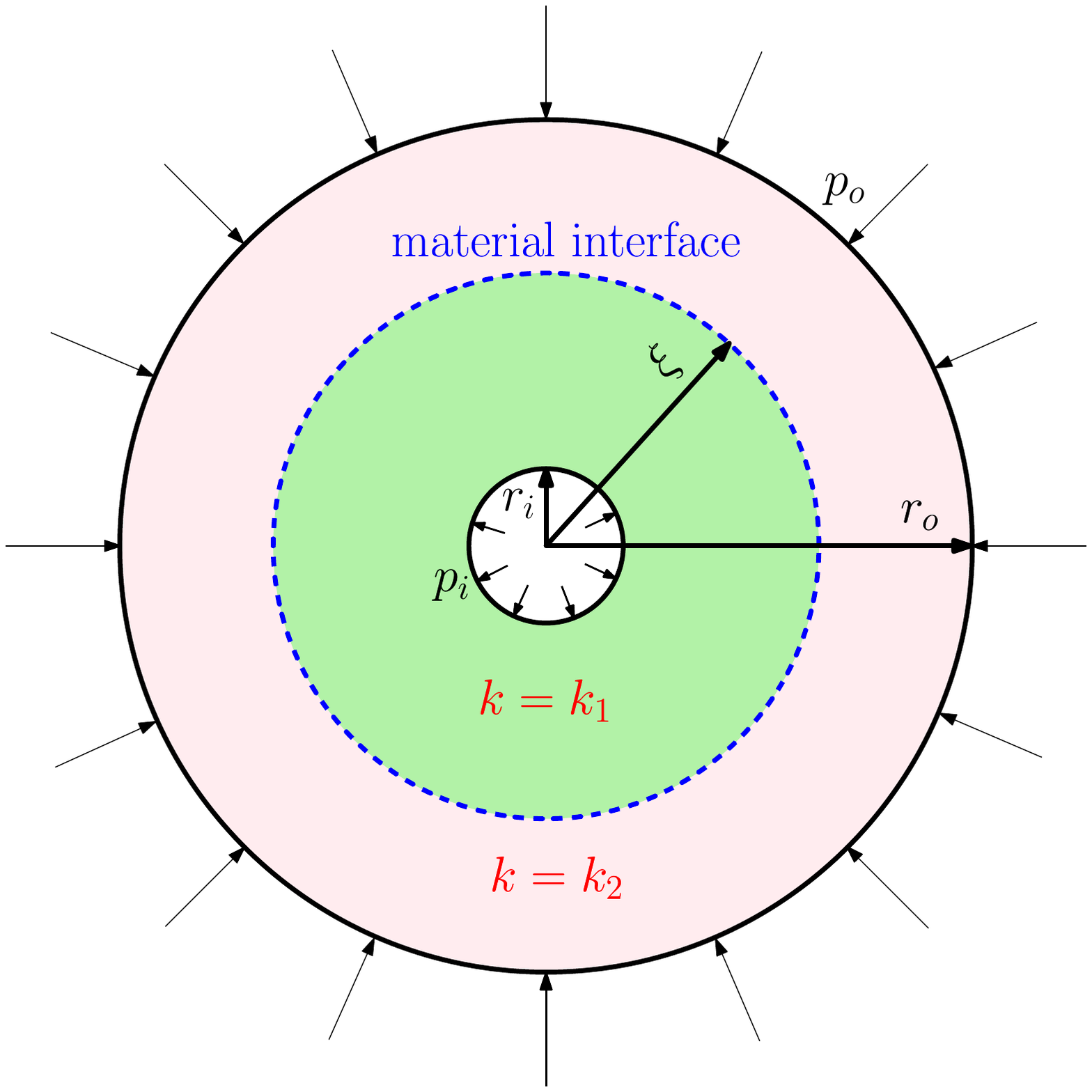}
	\caption{\textsf{2D axisymmetric problem:} The computational domain
	comprises two concentric circles. The inner and outer boundaries
	are subject to pressure loadings: $\widehat{\mathbf{n}} \cdot \mathbf{T} \widehat{\mathbf{n}} \vert_{r = r_i} = - p_i$ and $\widehat{\mathbf{n}} \cdot \mathbf{T} \widehat{\mathbf{n}} \vert_{r = r_o} = - p_o$. We have assumed a single material
	interface, as indicated in the figure, in deriving the optimal solution. \label{Fig:Conc_cyln_BVP}}
\end{figure}

The problem considered in this section is similar to the one solved by \cite{phatak2020optimal} for the case of the Darcy model. But herein we will extend the material design to the Darcy-Brinkman model.  The computational domain is sandwiched between two concentric circles with radii $r_i$ and $r_o > r_i$. The inner and outer boundaries are subject to pressure loadings $p_i$ and $p_o < p_i$, respectively; see figure \ref{Fig:Conc_cyln_BVP} for a pictorial description. The body force is neglected. \emph{The design problem is to find the optimum material distribution of two porous materials, one with low permeability, $k_L$, and other with higher permeability, $k_H > k_L$, within a circular domain.}


\subsection{Analytical solution for the optimal design}
We assume that each material exists in a symmetric, contiguous manner. We also assume that a single boundary exists between the two materials over the entire domain (cf. material interface shown in figure \ref{Fig:Conc_cyln_BVP}); this assumption will be validated using numerical simulations. To derive the analytical solution, we will exploit axisymmetry and use cylindrical polar coordinates $(r, \theta)$. The radial and tangential unit vectors are, respectively, denoted by
$\widehat{\mathbf{e}}_{r}$ and $\widehat{\mathbf{e}}_{\theta}$.

The assumptions mentioned above will allow us to write the spatial dependence of permeability field as follows:
\begin{align}
  \label{Eqn:Brinkman_2Daxi_k_field}
  k(r) = \left\{\begin{array}{ll}
  k_1 & r_i < r < \xi \\
  k_2 & \xi < r < r_o
  \end{array} \right.
\end{align}
Noting the inherent radial symmetry in the problem, we will consider only two-dimensional radial flows; the tangential component of the velocity is zero (i.e., $v_{\theta} = 0$). Thus, the velocity vector field takes the following form:
\begin{align}
  \mathbf{v(x)} = v_r (r) \; \widehat{\mathbf{e}}_r
\end{align}
where $v_r(r)$ denotes the radial component of the velocity. Likewise, we will assume the following functional form for the pressure field:
\begin{align}
  p(\mathbf{x}) = p(r)
\end{align}

The governing equations corresponding to the primal analysis under the Darcy-Brinkman model take the following form:
\begin{subequations}
  \begin{alignat}{2}
    \label{Eqn:TO_porous_Cylinder_DB}
    &\frac{\mu}{k(r)} v_r + \frac{d p}{d r} - 2 \mu \frac{1}{r} \frac{d}{dr} \left( r \frac{dv_r}{dr} \right) = 0
    &&\quad \forall r \in (r_i, r_o) \\
    \label{Eqn:TO_porous_Cylinder_Continuity}
    &\frac{1}{r} \frac{d (r v_r)}{d r} = 0
    &&\quad \forall r \in \; (r_i,r_o) \\
    \label{Eqn:TO_porous_Cylinder_BCi}
    -&p + 2 \mu \frac{dv_r}{dr}
    = - p_{i}
    &&\quad \mathrm{at} \; r = r_i \\
    \label{Eqn:TO_porous_Cylinder_BCo}
    -&p + 2 \mu \frac{dv_r}{dr}
    = - p_{\mathrm{o}}
    &&\quad \mathrm{at} \; r = r_o
  \end{alignat}
\end{subequations}
The jump conditions at the material interface take the following form:
\begin{subequations}
\begin{align}
    \label{Eqn:TO_porous_Cylinder_jump_velocity}
    &v_{r}(r \rightarrow \xi^{-}) = v_r(r \rightarrow \xi^{+}) \\
    \label{Eqn:TO_porous_Cylinder_jump_traction}
    &\left[-p + 2 \mu \frac{dv_r}{dr} \right]_{r \rightarrow \xi^{-}}
    = \left[-p + 2 \mu \frac{dv_r}{dr} \right]_{r \rightarrow \xi^{+}}
\end{align}
\end{subequations}
where $\xi^{\pm}$ denotes the one-sided limits. The jump conditions, respectively, imply that the radial component of the velocity and traction are continuous at the material interface.

We will start the derivation of the analytical solution by expanding the continuity equation \eqref{Eqn:TO_porous_Cylinder_Continuity} as follows:
\begin{align}
  \frac{1}{r} \frac{d (r v_r)}{d r}
  = \frac{d v_r}{dr} + \frac{v_r}{r} = 0
\end{align}
The above equation along with the first jump condition \eqref{Eqn:TO_porous_Cylinder_jump_velocity} imply that
\begin{align}
  \left. \frac{d v_r}{dr} \right|_{r \rightarrow \xi^{-}} =
    \left. \frac{d v_r}{dr} \right|_{r \rightarrow \xi^{+}}
\end{align}
The above equation implies that the second jump condition is equivalent to the pressure being continuous at the material interface:
\begin{align}
  \label{Eqn:TO_porous_Cylinder_jump_pressure}
  &p(r \rightarrow \xi^{-}) = p(r \rightarrow \xi^{+})
\end{align}

Equations \eqref{Eqn:TO_porous_Cylinder_Continuity} and \eqref{Eqn:TO_porous_Cylinder_jump_velocity} imply that the radial component of the velocity can be written as follows:
\begin{align}
  \label{Eqn:TO_porous_Cylinder_Vel}
  v_r(r) = \frac{A}{r} \quad \forall r \in (r_i, r_o)
\end{align}
where $A$ is a constant to be determined from the boundary conditions. Given this expression for $v_r$, equation \eqref{Eqn:Brinkman_2Daxi_k_field} and \eqref{Eqn:TO_porous_Cylinder_DB} imply the mathematical representation for the pressure field:
\begin{align}
  p(r) = \left\{\begin{array}{ll}
  -\frac{\mu}{k_1} A \ln(r) + B_1 & \quad r_i \leq r < \xi \\ \\
  -\frac{\mu}{k_2} A \ln(r) + B_2 & \quad \xi < r \leq r_o
  \end{array}\right.
\end{align}
where $B_1$ and $B_2$ are constants.

Enforcing the normal component of traction boundary condition based on the prescribed ambient pressure at the boundaries, equations \eqref{Eqn:TO_porous_Cylinder_BCi} and \eqref{Eqn:TO_porous_Cylinder_BCo} give us :
\begin{align}
  \label{Eqn:TO_porous_Cylinder_TractionBC1}
  \frac{\mu}{k_1} A \ln(r_i) -B_1 - \frac{2 \mu A}{r_i^2} = -p_i \\
  \label{Eqn:TO_porous_Cylinder_TractionBC2}
  \frac{\mu}{k_2} A \ln(r_o) -B_2 - \frac{2 \mu A}{r_o^2} = -p_o
\end{align}
The jump condition for the pressure \eqref{Eqn:TO_porous_Cylinder_jump_pressure} implies the following:
\begin{align}
  \label{Eqn:TO_porous_Cylinder_PressureBC1}
  B_1 - B_2 = A \; \mu \; \ln(\xi) \left( \frac{1}{k_1} - \frac{1}{k_2} \right)
\end{align}
By solving the above three equations \eqref{Eqn:TO_porous_Cylinder_TractionBC1}--\eqref{Eqn:TO_porous_Cylinder_PressureBC1}, we get the following expressions:
\begin{align}
  A &= \frac{(p_i - p_o)}{\mu}  \Upsilon^{-1}_{\mathrm{2D,DB}}(\xi)  \\
  B_1 &= p_i + \mu A \left[ \frac{1}{k_1} \ln {r_i}  - \frac{2}{r_i^2} \right] \\
  B_2 &= p_o + \mu A \left[ \frac{1}{k_2} \ln {r_o}  - \frac{2}{r_o^2} \right]
\end{align}
where $\Upsilon_{\mathrm{2D,DB}}(\xi)$, introduced for convenience, is defined as follows:
\begin{align}
  \label{Eqn:TopOpt_Upsilon_2D}
  \Upsilon_{\mathrm{2D,DB}}(\xi) := \frac{1}{k_1}
  \ln\left(\frac{\xi}{r_i}
  \right) + \frac{1}{k_2}
  \ln \left(\frac{r_o}{\xi}\right)
  + 2 \left(\frac{1}{r_i^2} - \frac{1}{r_o^2}\right)
\end{align}
Accordingly, the radial velocity takes the following form:
\begin{align}
  v(r) = \frac{(p_i - p_o)}{\mu \, r} \Upsilon_{\mathrm{2D,DB}}^{-1}(\xi)
\end{align}
Based on the above solution, the total rate of dissipation takes the following form:
\begin{alignat}{1}
  \Phi(\xi) &= \int_{r_i}^{r_o} \frac{\mu}{k} \left(v_r\right)^2 (2 \pi r) dr
  + \int_{r_i}^{r_o} 2 \mu \left(\left(\frac{d v_r}{d r}\right)^2
  + \left(\frac{v_r}{r}\right)^2\right)(2 \pi r) dr \notag \\
  &= 2 \pi \mu A^2 \left(\frac{1}{k_1} \int_{r_i}^{\xi^{-}} \frac{1}{r} dr + \frac{1}{k_2} \int_{\xi^{+}}^{r_o} \frac{1}{r} dr
  \right) + 8 \pi \mu A^2 \left( \int_{r_i}^{r_o} \frac{1}{r^3} dr \right)                                                       \notag  \\
  &=  \frac{2 \pi (p_i - p_o)^2}{\mu} \Upsilon^{-1}_{\mathrm{2D,DB}}(\xi)
\end{alignat}

For the same boundary value problem, the corresponding expression for the total rate of dissipation under the Darcy model is:
\begin{alignat}{1}
  \Phi_{\mathrm{D}}(\xi) &= \frac{2 \pi (p_i - p_o)^2}{\mu} \Upsilon^{-1}_{\mathrm{2D,D}}(\xi)
\end{alignat}
where 
\begin{align}
  \Upsilon_{\mathrm{2D,D}}(\xi) := \frac{1}{k_1}
  \ln\left(\frac{\xi}{r_i}
  \right) + \frac{1}{k_2}
  \ln \left(\frac{r_o}{\xi}\right)
\end{align}
Since $r_i$ and $r_o$ are given constants and are independent of the design variable $\xi$, $\Upsilon_{\mathrm{2D,DB}}(\xi) \propto \Upsilon_{\mathrm{2D,D}}(\xi)$. Moreover, $\Upsilon_{\mathrm{2D,DB}}(\xi)$ is a convex function of $\xi$, as $\Upsilon_{\mathrm{2D,D}}(\xi)$ is shown to be a convex function \citep{phatak2020optimal}. (For a definition of a convex function, see \citep{boyd2004convex}.)

Noting that $\Phi(\xi) \propto \Upsilon^{-1}_{\mathrm{2D,DB}}(\xi)$, the material design problem becomes:
\begin{align}
  \label{Eqn:TopOpt_2D_axisym_optim_problem}
  \widehat{\xi}_{\mathrm{2D}} \leftarrow
  \left\{
  \begin{array}{l}
    \mathop{\mathrm{argmax}}_{\xi} \; \Phi(\xi) \equiv
    \mathop{\mathrm{argmin}}_{\xi} \; \Upsilon_{\mathrm{2D,DB}}(\xi)
    \equiv \mathop{\mathrm{argmin}}_{\xi} \; \Upsilon_{\mathrm{2D,D}}(\xi) \\
    \\ 
    \mbox{subject to} \quad
    \frac{2 \pi (\xi^2 - r_i^2)}{2 \pi (r_o^2 - r_i^2)} \leq \gamma 
    \quad \mbox{(volume constraint)}
  \end{array}
 \right.
\end{align}
where $\widehat{\xi}_{\mathrm{2D}}$ is the optimal location of material interface, and $0 \leq \gamma \leq 1$ is the user-specified volumetric bound placed on the constrained material (i.e., in this case, the high-permeability material). The above constrained optimization problem \eqref{Eqn:TopOpt_2D_axisym_optim_problem} reveals that the optimal material distribution is identical under the Darcy and Darcy-Brinkman models.

Given the single material interface assumption, the design problem reduces to identifying the case among two possibilities that minimizes $\Upsilon_{\mathrm{2D}}(\xi)$ while meeting the volume constraint; the two possibilities are whether to place the high-permeability material near the inlet and the low-permeability material near the outlet, or \emph{vice versa}. This reduced problem is identical to the one solved in \citep{phatak2020optimal} for the case of the Darcy model. Following the same reasoning used for the Darcy model, the optimum design even for the Darcy-Brinkman model is to place the high(low)-permeability material near the inlet (outlet), and the location of the material interface is:  
\begin{align}
  \label{Eqn:TopOpt_2D_axisym_xiopt}
  \widehat{\xi}_{\mathrm{2D}} = \sqrt{(1 - \gamma) r_{i}^2 + \gamma r_o^2}
\end{align}
The corresponding maximum rate of dissipation under these two models are:
\begin{alignat}{2}
  \max_{\xi} \; \Phi =
  \left\{\begin{array}{ll}
  \Phi_\mathrm{D}(\widehat{\xi})
  = \frac{2 \pi (p_i - p_o)^2}{\mu}
  \Upsilon^{-1}_{\mathrm{2D,D}}(\widehat{\xi}_{\mathrm{2D}})
  & \quad \mbox{Darcy model} \\
  \\
  \Phi_\mathrm{DB}(\widehat{\xi}) = \frac{2 \pi (p_i - p_o)^2}{\mu}
  \Upsilon^{-1}_{\mathrm{2D,DB}}(\widehat{\xi}_{\mathrm{2D}})
  & \quad \mbox{Darcy-Brinkman model}
	\end{array} \right.
\end{alignat}

Table \ref{Table:Brinkman_2Daxi_Comparison} summarizes the optimal material distribution and the solution fields at the optimal design under the Darcy-Brinkman and Darcy models.

\begin{table}
  \renewcommand{\arraystretch}{1.5} 
  \caption{\textsf{2D axisymmetric problem:} A comparison of material designs and solution fields at the optimal design under the Darcy-Brinkman and Darcy models. \label{Table:Brinkman_2Daxi_Comparison}}
  \begin{tabular}{|c|c|c|}\hline
    \textbf{Quantity} & \textbf{Darcy-Brinkman model} & \textbf{Darcy model} \\ \hline
    \multicolumn{3}{|c|}{\emph{Optimal location of the material interface}} \\ \hline
    $\widehat{\xi}_{\mathrm{2D}}$ & $\widehat{\xi}_{\mathrm{2D,DB}} = \widehat{\xi}_{\mathrm{2D,D}}$ & $\widehat{\xi}_{\mathrm{2D,D}} = \sqrt{(1 - \gamma) r_{i}^2 + \gamma r_o^2}$ \\ \hline
    \multicolumn{3}{|c|}{\emph{Solution fields under the optimal material distribution}} \\ \hline
    $\Upsilon_{\mathrm{2D}}(\xi)$ & $\Upsilon_{\mathrm{2D,DB}}(\xi) = \Upsilon_{\mathrm{2D,D}}(\xi) + 2 \left( \frac{1}{r_i^2} - \frac{1}{r_o^2} \right)$ & $\Upsilon_{\mathrm{2D,D}}(\xi) =  \frac{1}{k_1}
    \ln\left(\frac{\xi}{r_i}
    \right) + \frac{1}{k_2}
    \ln \left(\frac{r_o}{\xi}\right)$ \\
    $v_r(r_i \leq r \leq r_o)$ & $\frac{(p_i - p_o)}{\mu \, r} \Upsilon^{-1}_{\mathrm{2D,DB}}(\widehat{\xi})$ & $\frac{(p_i - p_o)}{\mu \, r} \Upsilon^{-1}_{\mathrm{2D,D}}(\widehat{\xi})$ \\ 
    $p(r_i \leq r \leq \widehat{\xi})$ & $p_i + (p_i - p_o) \Upsilon^{-1}_{\mathrm{2D,DB}}(\xi) \left[ \frac{1}{k_1} \ln \left( \frac{r_i}{r} \right) - \frac{2}{r_i^2} \right]$ & $p_i + (p_i - p_o) \Upsilon^{-1}_{\mathrm{2D,D}}(\xi) \left[ \frac{1}{k_1} \ln \left( \frac{r_i}{r} \right)  \right]$ \\
    $p(\widehat{\xi} \leq r \leq r_o) $ & $p_o + (p_i - p_o) \Upsilon^{-1}_{\mathrm{2D,DB}}(\xi) \left[ \frac{1}{k_2} \ln \left( \frac{r_o}{r} \right) - \frac{2}{r_o^2} \right]$ & $p_o + (p_i - p_o) \Upsilon^{-1}_{\mathrm{2D,D}}(\xi) \left[ \frac{1}{k_2} \ln \left( \frac{r_o}{r} \right)  \right]$ \\ \hline
    \multicolumn{3}{|c|}{\emph{Pressures within the domain at the inlet and outlet}} \\ \hline
	$p(r = r_i)$ & $p_i - \frac{2}{r_i^2} (p_i - p_o) \Upsilon^{-1}_{\mathrm{2D,DB}}(\xi)$ & $p_i$  \\
    $p(r = r_o)$ & $p_o - \frac{2}{r_o^2} (p_i - p_o) \Upsilon^{-1}_{\mathrm{2D,DB}}(\xi)$ & $p_o$  \\ \hline
    \multicolumn{3}{|c|}{\emph{Rate of dissipation for optimal material distribution}} \\ \hline
    $\Phi(\widehat{\xi})$ & $\frac{2 \pi (p_i - p_o)^2 \Upsilon^{-1}_{\mathrm{2D,DB}}(\widehat{\xi})}{\mu}$ & $\frac{2 \pi (p_i - p_o)^2 \Upsilon^{-1}_{\mathrm{2D,D}}(\widehat{\xi})}{\mu}$ \\ \hline
  \end{tabular}
\end{table}

\subsection{Numerical solution for optimal material layouts}
\begin{figure}[h]
	\subfigure{\includegraphics[scale=0.25]{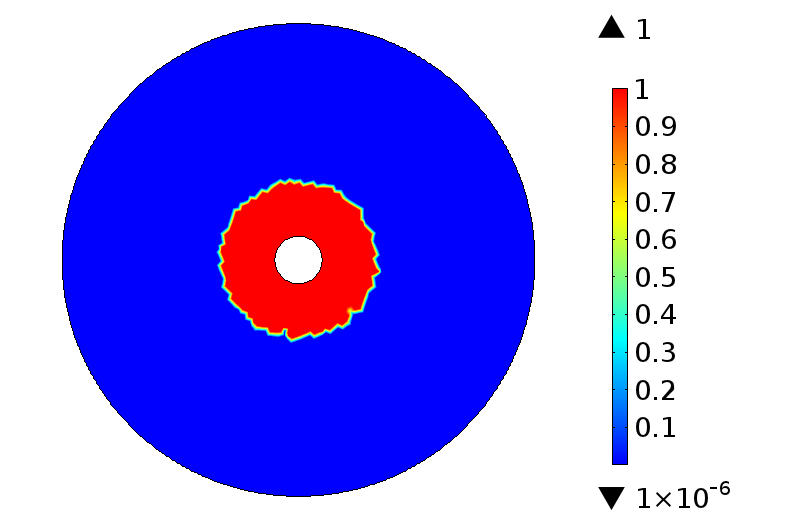}}
	\subfigure{\includegraphics[scale=0.25]{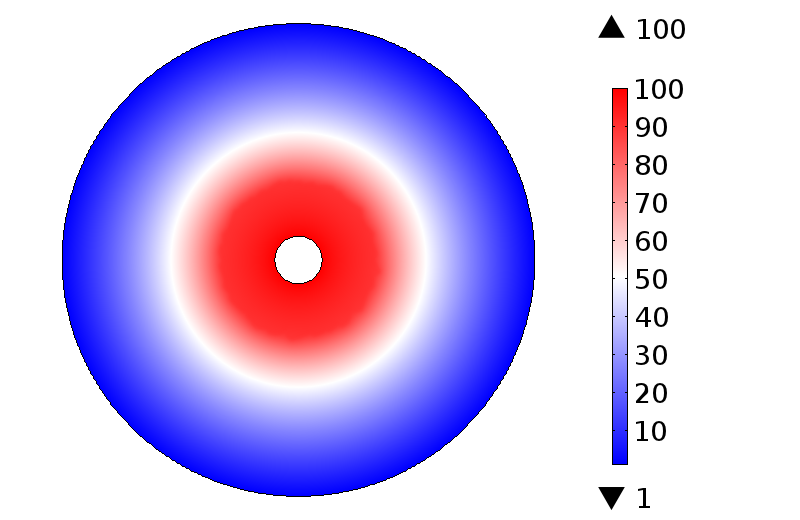}}
	\subfigure{\includegraphics[scale=0.25]{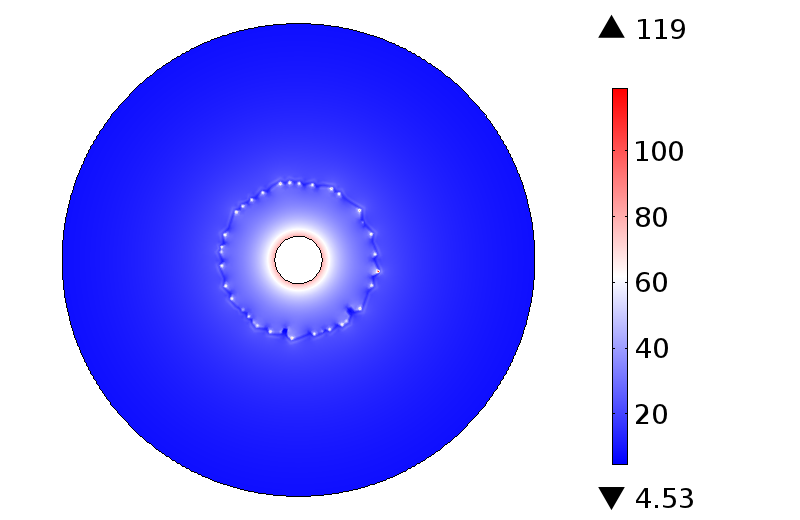}}

	{\small (i) Darcy model.}

	\subfigure{\includegraphics[scale=0.25]{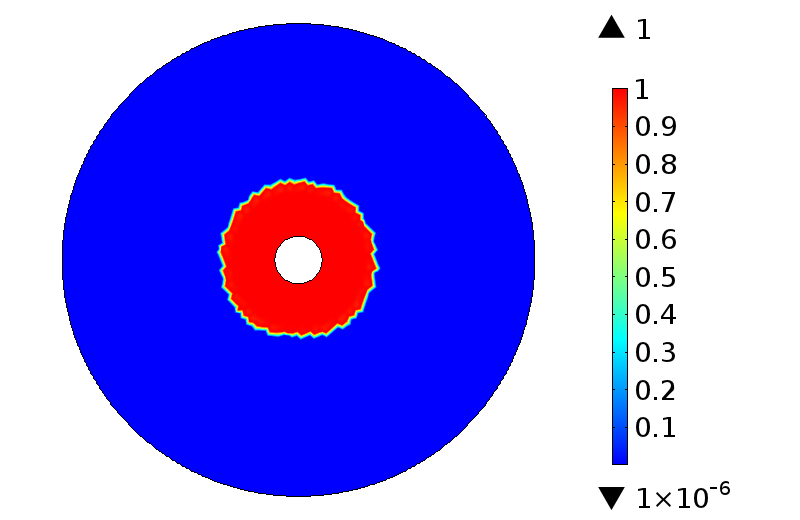}}
	\subfigure{\includegraphics[scale=0.25]{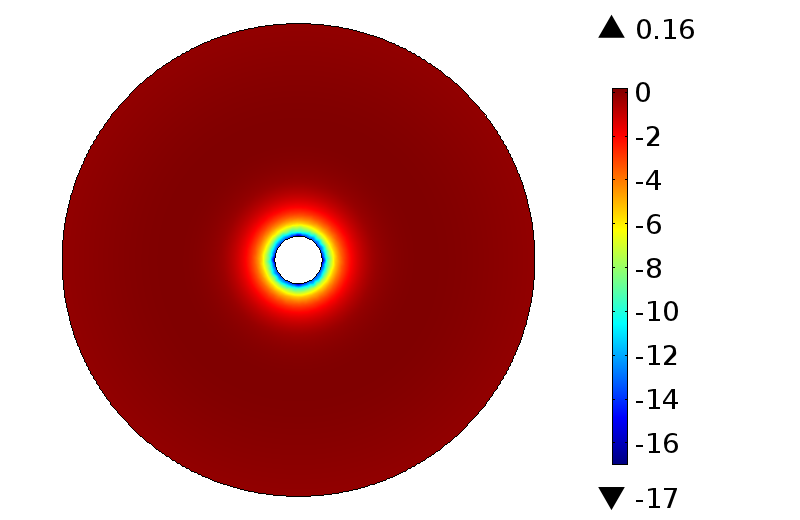}}
	\subfigure{\includegraphics[scale=0.25]{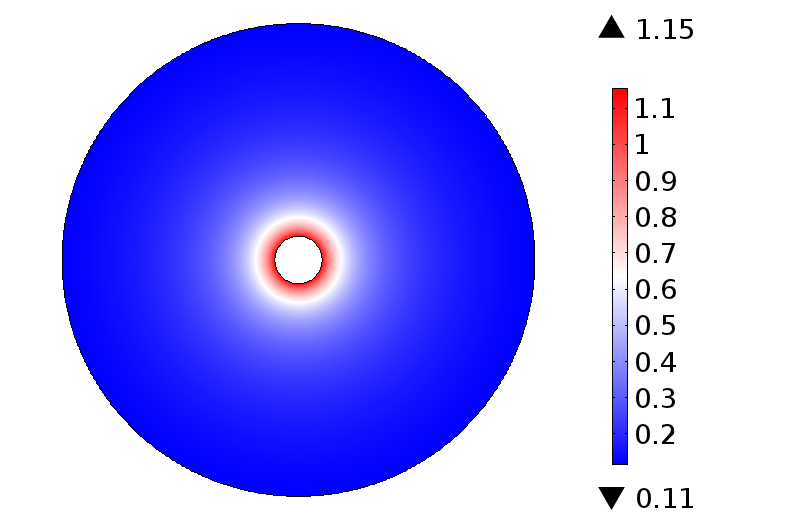}}

	{\small (ii) Darcy-Brinkman model.}

	\caption{\textsf{2D axisymmetric problem:} This figure shows the optimal material distribution (left panel) and the corresponding pressure profile (middle) and velocity profile (right) under the Darcy and Darcy-Brinkman models. The domain is subject to zero body force as well as zero mass source. The limiting area of the constrained material (herein, the high-permeability material) is $\gamma = 0.1$ times the total area of the domain. On the left panel, the regions occupied by the constrained material are indicated in `red' while `blue' represents the regions occupied by the unconstrained material (i.e., the low-permeability material).  (See the online version for the figure in color.) 
          \label{Fig:2D_Dist1}}
\end{figure}

The parameters used in the numerical simulation are provided in Table \ref{Fig:2D_Circles_BVP_parameters}. Figure \ref{Fig:2D_Dist1} shows the material distribution, the pressure and velocity profiles within the domain under the Darcy and Darcy-Brinkman models. The main conclusions from this figure are:
\begin{enumerate}
	\item The numerical solution matches with the analytical solution.
	\item Numerical solutions justify the single interface assumption, which is invoked for deriving the analytical, even for the Darcy-Brinkman model. A similar assumption is shown to be valid for the Darcy model, again using numerical simulations, in our previous paper \citep{phatak2020optimal}.
	\item The optimal material design under the Darcy-Brinkman model is exactly the same as that of the Darcy model. This similarity is due underlying axisymmetry in the problem which causes the viscous shearing stress to vanish.
	\item However, there are marked differences between the solutions for the two models. Under the Darcy-Brinkman model, the pressures at the inlet and outlet are, respectively, lower than the prescribed pressure loadings at these locations. On the other hand, under the Darcy model, the pressure at the boundary is equal to the prescribed boundary pressure. Mathematically, pressures under the Darcy model satisfy the maximum-minimum principle---the maximum and minimum pressures occur on the boundary \citep{shabouei_nakshatrala_cicp}. A similar principle is not available for pressures under the Darcy-Brinkman model.
\end{enumerate}

\begin{table}[h]
  \caption{This table provides the parameters used in the numerical simulations
    of the 2D and 3D axisymmetric problems.
    \label{Fig:2D_Circles_BVP_parameters}}
  \begin{tabular}{|lr||lr|}\hline
    parameter & value & parameter & value \\ \hline
    $r_i$ & 0.1 & $r_o$ & 1.0 \\
    $p_i$ & 100 & $p_o$ & 1 \\
    $k_L$ & 0.1 & $k_H$ & 1 \\
    $\gamma$ & 0.1 & $\mu$ & 1 \\ \hline
  \end{tabular}
\end{table}

\section{OPTIMAL LAYOUTS FOR A 3D AXISYMMETRIC PROBLEM}
\label{Sec:S5_Brinkman_AxiSym_3D}
\begin{figure}[h]
  \includegraphics[scale=1.25]{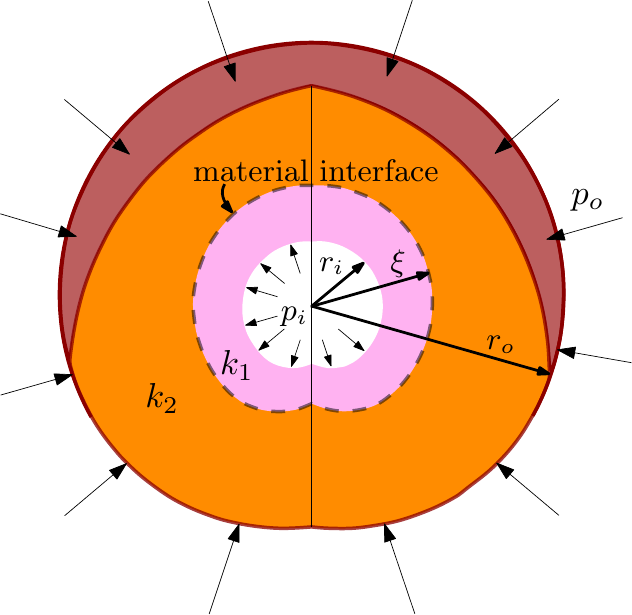}
  \caption{\textsf{3D axisymmetric problem:} The computational domain comprises two concentric spheres. The inner and outer boundaries are subject to pressure loadings: $\widehat{\mathbf{n}} \cdot \mathbf{T} \widehat{\mathbf{n}} \vert_{r = r_i} = - p_i$ and $\widehat{\mathbf{n}} \cdot \mathbf{T} \widehat{\mathbf{n}} \vert_{r = r_o} = - p_o$. We have assumed a single material interface, as indicated in the figure, in deriving the optimal solution. \label{Fig:Conc_sphere_BVP}}
\end{figure}

The three-dimensional axisymmetric problem is an extension of the two-dimensional problem with similar pressure/traction boundary conditions. Figure \ref{Fig:Conc_sphere_BVP} provides a pictorial description of the problem.

The computational domain comprises two concentric spheres with an inner sphere of radius $r_i$ and an outer sphere of radius $r_o > r_i$. For the Darcy-Brinkman model, traction $\mathrm{\mathbf{t}^p(\mathbf{x})} = -p_i \mathrm{\widehat{\mathbf{n}}}(\mathbf{x})$ is prescribed at the inner boundary and $\mathrm{\mathbf{t}^p(\mathbf{x})} = -p_o \mathrm{\widehat{\mathbf{n}}}(\mathbf{x})$ at the outer boundary. Correspondingly, for the Darcy model, pressure $p_i$ is applied at the inner boundary while pressure $p_o < p_i$ is applied to the outer boundary. The body force is neglected. The parameters used in this problem are provided in Table \ref{Fig:2D_Circles_BVP_parameters}. The objective function is to maximize the rate of dissipation with a volume constraint placed on the high permeability material, specified via a user-specified value for $\gamma$. Axisymmetry is invoked for both the models.

\subsection{Analytical solution for the optimal design}
We will consider the spherical polar coordinates to make use of symmetry:
\begin{align}
  r_i \leq r \leq r_o, \quad
  0 \leq \theta \leq \pi, \quad 
  0 \leq \phi \leq 2\pi 
\end{align}
where $r$ is the radial distance from the center, and $\theta$ and $\phi$ are the azimuth and polar angles, respectively. We assume that each material is present in a symmetric, contiguous manner and a single boundary exists between the two materials over the entire domain. We can, therefore, state permeability as:
\begin{align}
  \label{Eqn:Brinkman_permeability_field}
  k(r) = \left\{\begin{array}{ll}
  k_1 & r_i < r < \xi \\
  k_2 & \xi < r < r_o
  \end{array} \right.
\end{align}
where $r = \xi$ is the location of the material interface. The underlying symmetry of the problem enables us to represent the solution fields as follows: 
\begin{align}
  \mathbf{v}(\mathbf{x}) = v_{r}(r) \, \hat{\mathbf{e}}_{r}
  \quad \mathrm{and} \quad
  p(\mathbf{x}) = p(r) 
\end{align}
where $v_{r}$ is radial component of the velocity, and $\hat{\mathbf{e}}_r$ is the unit vector along the radial direction.

The governing equations for the primal analysis under the Darcy-Brinkman model can be written as follows:
\begin{subequations}
  \begin{alignat}{2}
    \label{Eqn:DB_porous_Sphere_BLM}
    &\frac{1}{{k}({r})} {v}_r
    + \frac{dp}{dr} - 2 {\mu} \left[ \frac{1}{{r}^2} \frac{d}{d{r}} (r^2 \frac{dv_r}{dr}) - \frac{2v_r}{r^2} \right]  = 0
    &&\quad \forall {r} \in (r_i, r_o) \\
    \label{Eqn:DB_porous_Sphere_Continuity}
    &\frac{1}{r^2} \frac{d ({r}^2 {v}_r)}{dr} = 0
    &&\quad \forall {r} \in ({r}_i, r_o) \\
    \label{Eqn:DB_porous_Sphere_BCi}
    -&p + 2 \mu \frac{dv_r}{dr} = - p_{\mathrm{i}}
    &&\quad \mathrm{at} \, r = r_i \\
    \label{Eqn:DB_porous_Sphere_BCo}
    -&p + 2 \mu \frac{dv_r}{dr} = - p_{\mathrm{o}}
    &&\quad \mathrm{at} \, r = r_o
  \end{alignat}
\end{subequations}
The jump conditions along the interface take the following form:
\begin{subequations}
  \begin{align}
    \label{Eqn:TopOpt_3D_velocity_jump}
    &v_r(r \rightarrow \xi^{-}) = v_r(r \rightarrow \xi^{+}) \\
    \label{Eqn:TopOpt_3D_traction_jump}
    &\left[-p + 2 \mu \frac{dv_r}{dr}\right]_{r \rightarrow \xi^{-}}
    = \left[-p + 2 \mu \frac{dv_r}{dr}\right]_{r \rightarrow \xi^{+}}
  \end{align}
\end{subequations}

Following the procedure using in the previous section, we expand the continuity equation \eqref{Eqn:DB_porous_Sphere_Continuity} and note that $r \neq 0$:
\begin{align}
  \frac{dv_r}{dr} + \frac{2 v_r}{r} = 0 
\end{align}
Using the continuity of the radial component of the velocity across the material interface, we conclude that
\begin{align}
  \left[\frac{dv_r}{dr}\right]_{r \rightarrow \xi^{-}} =
    \left[\frac{dv_r}{dr}\right]_{r \rightarrow \xi^{+}} 
\end{align}
The above equation along with the continuity of tractions across $r = \xi$ imply the continuity of the pressure field across the material interface:  
\begin{align}
  \label{Eqn:TopOpt_3D_pressure_jump}
  &p(r \rightarrow \xi^{-}) = p(r \rightarrow \xi^{+})
\end{align}

Noting that $r \neq 0$, equations \eqref{Eqn:DB_porous_Sphere_Continuity} and \eqref{Eqn:TopOpt_3D_velocity_jump} give us the following representation for the velocity field:
\begin{align}
  \label{Eqn:TO_porous_Sphere_Velocity}
  v(r) = \frac{A}{r^2} \quad r_i \leq r \leq r_o
\end{align}
where $A$ is a constant to be determined. Using equations \eqref{Eqn:DB_porous_Sphere_BLM}, \eqref{Eqn:TO_porous_Sphere_Velocity} and \eqref{Eqn:Brinkman_permeability_field}, the solution for the pressure field can be written as follows:
\begin{align}
  p(r) = \left\{ \begin{array}{cc} 
    \frac{\mu}{k_1 r} A + B_1 & \quad r_i \leq r < \xi \\ \\
    \frac{\mu}{k_2 r} A + B_2 & \quad \xi < r \leq r_o
  \end{array} \right.
\end{align}
where $B_1$ and $B_2$ are constants. Traction boundary conditions \eqref{Eqn:DB_porous_Sphere_BCi} and \eqref{Eqn:DB_porous_Sphere_BCo} imply that  
\begin{align}
  \label{Eqn:TO_porous_Sphere_BC1}
  - \frac{\mu}{k_1} \frac{A}{r_i} - B_1 - 2 \mu \left( \frac{2A}{r_i^3} \right) = -p_i  \\
  \label{Eqn:TP_porous_Sphere_BC2}
  - \frac{\mu}{k_2} \frac{A}{r_o} - B_2 - 2 \mu \left( \frac{2A}{r_o^3} \right) = -p_o	
\end{align}
The jump condition for the pressure field, given by equation \eqref{Eqn:TopOpt_3D_pressure_jump}, implies that 
\begin{align}
  \label{Eqn:TO_porous_Sphere_JC}
  \frac{\mu}{k_1} \frac{A}{\xi} + B_1 = \frac{\mu}{k_2} \frac{A}{\xi} + B_2
\end{align}
By solving the above three equations, we get the following expressions for the three constants:
\begin{align}
  A &= \frac{(p_i - p_o)}{\mu} \Upsilon_{\mathrm{3D,DB}}^{-1}(\xi) \\
  B_1 &= p_i - \mu A \left[\frac{1}{k_1 r_i} + \frac{4}{r_i^3} \right] \\
  B_2 &= p_o - \mu A \left[ - \frac{1}{k_2 r_o} + \frac{4}{r_o^3} \right] 
\end{align}
where
\begin{align}
  \Upsilon_{\mathrm{3D,DB}}(\xi)  = \left[ \frac{1}{k_1} \left( \frac{1}{r_i} - \frac{1}{\xi} \right) + \frac{1}{k_2} \left( \frac{1}{\xi} - \frac{1}{r_o} \right) \right] + 4 \left( \frac{1}{r_i^3} - \frac{1}{r_o^3} \right)
\end{align}

Thus, the total rate of dissipation under the Darcy-Brinkman model will be: 
\begin{align}
  \Phi(\xi) &= \int_{r_i}^{r_o} \frac{\mu}{k} \left(v_r\right)^2 (4 \pi r^2) dr + \int_{r_i}^{r_o} 2 \mu \left( \left(\frac{dv_r}{dr}\right)^2 + 2 \left(\frac{v_r}{r}\right)^2 \right) (4 \pi r^2) dr \notag \\
  &= 2 \pi \mu A^2 \left(\frac{1}{k_1} \int_{r_i}^{\xi^{-}} \frac{1}{r^2} dr + \frac{1}{k_2} \int_{\xi^{+}}^{r_o} \frac{1}{r^2} dr
  \right) + 48 \pi \mu A^2
  \left( \int_{r_i}^{r_o} \frac{1}{r^4} dr \right) \notag  \\
  &=  \frac{4 \pi (p_i - p_o)^2}{\mu} \Upsilon^{-1}_{\mathrm{3D,DB}}(\xi) 
\end{align}
For the same boundary value problem, the total rate of dissipation under the Darcy model is:
\begin{subequations}
  \begin{alignat}{1}
    \Phi_{\mathrm{D}}(\xi) &=
    \frac{4 \pi (p_i - p_o)^2}{\mu} \Upsilon^{-1}_{\mathrm{3D,D}}(\xi) 
  \end{alignat}
\end{subequations}
where
\begin{align}
    \Upsilon_{\mathrm{3D,D}}(\xi)  = \left[ \frac{1}{k_1} \left( \frac{1}{r_i} - \frac{1}{\xi} \right) + \frac{1}{k_2} \left( \frac{1}{\xi} - \frac{1}{r_o} \right) \right] 
\end{align}
Since $r_i$ and $r_o$ are given constants and are independent of the design variable $\xi$, $\Upsilon_{\mathrm{3D,DB}}(\xi) \propto \Upsilon_{\mathrm{3D,D}}(\xi)$. Noting that $\Phi(\xi) \propto \Upsilon^{-1}_{\mathrm{3D,DB}}(\xi)$, we can conclude that:
\begin{align}
	\label{Eqn:TopOpt_3D_axisym_optim_problem}
	\widehat{\xi}_{\mathrm{3D}} \leftarrow
        \left\{ \begin{array}{l}
	  \mathop{\mathrm{argmax}}_{\xi} \; \Phi(\xi) \equiv 
	  \mathop{\mathrm{argmin}}_{\xi} \; \Upsilon_{\mathrm{3D,DB}}(\xi)
          \equiv \mathop{\mathrm{argmin}}_{\xi} \; \Upsilon_{\mathrm{3D,D}}(\xi) \\
          \\
          \mbox{subject to} \quad \frac{4\pi(\xi^{3} - r_i^3)/3}{4\pi(r_o^3 - \xi^3)/3} \leq \gamma \quad \mbox{(volume constraint)}
        \end{array} \right.
\end{align}
Similar to the 2D axisymmetric problem, the above optimization problem reveals that the optimal location of the material interface is identical under the Darcy-Brinkman and Darcy models. Following the same procedure as was done for the case of the Darcy model (see \citep{phatak2020optimal}), we find the optimal location of the material interface to be:
\begin{align}
  \label{Eqn:TopOpt_3D_axisym_xiopt}
  \widehat{\xi}_{\mathrm{3D}} = \sqrt[3]{(1 - \gamma) r_{i}^3 + \gamma r_o^3}
\end{align}
The optimum design, again, is to place the high-permeability material near the inlet and the low-permeability material near the outlet. The corresponding maximum rate of dissipation for these two models are:
\begin{align}
  \max_{\xi} \; \Phi = \left\{ \begin{array}{ll}
    \Phi_\mathrm{D}(\widehat{\xi}_{\mathrm{3D}}) = \frac{4 \pi (p_i - p_o)^2}{\mu}
    \Upsilon^{-1}_{\mathrm{3D,D}}(\widehat{\xi}_{\mathrm{3D}})
    & \mbox{Darcy model} \\
    \\
    \Phi_\mathrm{DB}(\widehat{\xi}_{\mathrm{3D}}) = \frac{4 \pi (p_i - p_o)^2}{\mu}
    \Upsilon^{-1}_{\mathrm{3D,DB}}(\widehat{\xi}_{\mathrm{3D}})
    & \mbox{Darcy-Brinkman model}
  \end{array} \right.
\end{align}

Table \ref{Table:Brinkman_3Daxi_Comparison} summarizes the optimal material distribution and the solution field at the optimal design under the Darcy-Brinkman and Darcy models.

\begin{table}
	\renewcommand{\arraystretch}{1.5} 
	\caption{\textsf{3D axisymmetric problem:} A comparison of material designs and solution fields at the optimal design under the Darcy-Brinkman and Darcy models. \label{Table:Brinkman_3Daxi_Comparison}}
	\begin{tabular}{|c|c|c|}\hline
		\textbf{Quantity} & \textbf{Darcy-Brinkman model} & \textbf{Darcy model} \\ \hline
		\multicolumn{3}{|c|}{\emph{Optimal location of the material interface}} \\ \hline
		$\widehat{\xi}_{\mathrm{3D}}$ & $\widehat{\xi}_{\mathrm{3D,DB}} = \widehat{\xi}_{\mathrm{3D,D}}$ & $\widehat{\xi}_{\mathrm{3D,D}} = \sqrt[3]{(1 - \gamma) r_{i}^3 + \gamma r_o^3}$ \\ \hline
		\multicolumn{3}{|c|}{\emph{Solution fields under the optimal material distribution}} \\ \hline
		$\Upsilon_{\mathrm{3D}}$ & $\Upsilon_{\mathrm{3D,DB}}(\xi) = \Upsilon_{\mathrm{3D,D}}(\xi)  + 4 \left( \frac{1}{r_i^3} - \frac{1}{r_o^3} \right)$ & $	\Upsilon_{\mathrm{3D,D}}(\xi)  = \left[ \frac{1}{k_1} \left( \frac{1}{r_i} - \frac{1}{\xi} \right) + \frac{1}{k_2} \left( \frac{1}{\xi} - \frac{1}{r_o} \right) \right]  $ \\ 
		$v_r(r_i \leq r \leq r_o)$ & $\frac{(p_i - p_o)}{\mu \, r^2} \Upsilon^{-1}_{\mathrm{3D,DB}}(\widehat{\xi})$ & $\frac{(p_i - p_o)}{\mu \, r^2} \Upsilon^{-1}_{\mathrm{3D,D}}(\widehat{\xi})$ \\ 
		$p(r_i \leq r \leq \widehat{\xi})$ & $p_i + (p_i - p_o) \Upsilon^{-1}_{\mathrm{3D,DB}}(\xi) \left[ \frac{1}{k_1 \, r}- \frac{1}{k_1 \, r_i}  - \frac{4}{r_i^3} \right] $ & $p_i + (p_i - p_o) \Upsilon^{-1}_{\mathrm{3D,D}}(\xi) \left[ \frac{1}{k_1 \, r}- \frac{1}{k_1 \, r_i} \right] $ \\
		$p(\widehat{\xi} \leq r \leq r_o) $ & $p_o + (p_i - p_o) \Upsilon^{-1}_{\mathrm{3D,DB}}(\xi)  \left[ \frac{1}{k_2 \, r} - \frac{1}{k_2 \, r_o}  - \frac{4}{r_o^3} \right] $ & $p_o + (p_i - p_o) \Upsilon^{-1}_{\mathrm{3D,D}}(\xi)  \left[ \frac{1}{k_2 \, r} - \frac{1}{k_2 \, r_o} \right] $  \\ \hline
		\multicolumn{3}{|c|}{\emph{Pressures within the domain at the inlet and outlet}} \\ \hline
		$p(r = r_i)$ & $p_i - \frac{4}{r_i^3} (p_i - p_o) \Upsilon^{-1}_{\mathrm{3D,DB}}(\xi)$ & $p_i$  \\
		$p(r = r_o)$ & $p_o - \frac{4}{r_o^3} (p_i - p_o) \Upsilon^{-1}_{\mathrm{3D,DB}}(\xi)$ & $p_o$  \\ \hline
		\multicolumn{3}{|c|}{\emph{Rate of dissipation for optimal material distribution}} \\ \hline
		$\Phi(\widehat{\xi})$ & $\frac{4 \pi (p_i - p_o)^2 \Upsilon^{-1}_{\mathrm{3D,DB}}(\widehat{\xi})}{\mu}$ & $\frac{4 \pi (p_i - p_o)^2 \Upsilon^{-1}_{\mathrm{3D,D}}(\widehat{\xi})}{\mu}$ \\ \hline
	\end{tabular}
\end{table} 

\subsection{Numerical solution for optimal material layouts}
\begin{figure}[h]
  \subfigure{\includegraphics[scale=0.25]{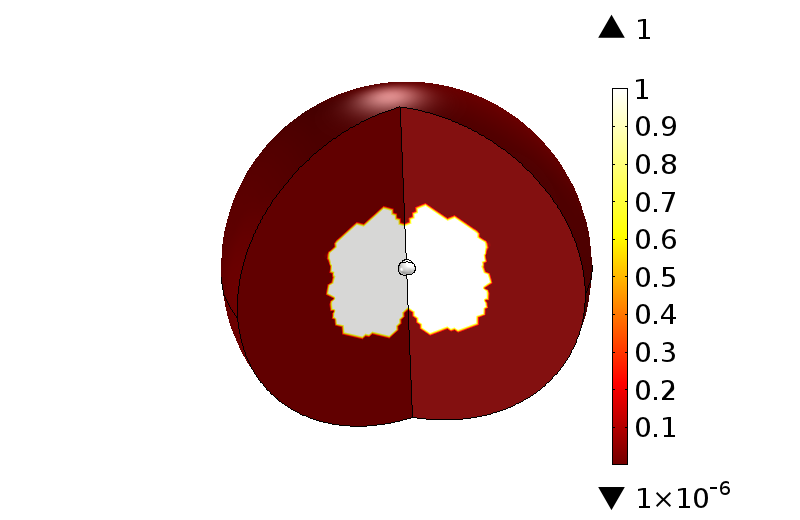}}
  \subfigure{\includegraphics[scale=0.25]{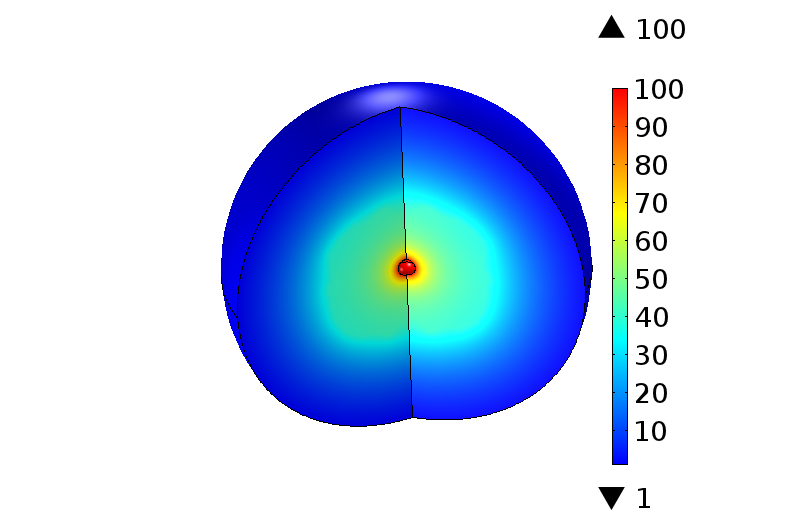}}
  \subfigure{\includegraphics[scale=0.25]{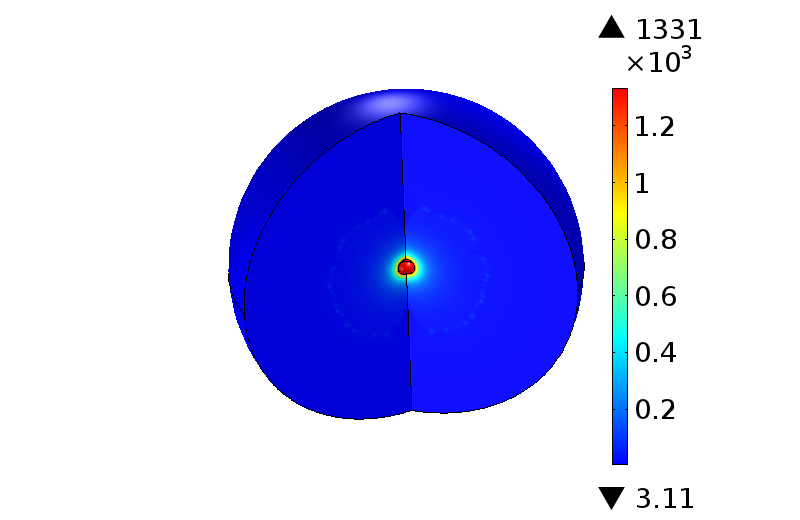}}
  
  {\small (i) Darcy model.}
  
  \subfigure{\includegraphics[scale=0.25]{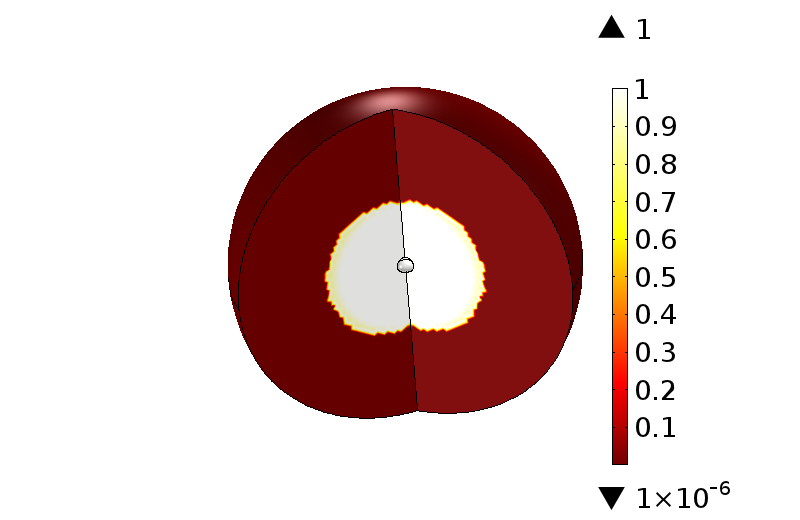}}
  \subfigure{\includegraphics[scale=0.25]{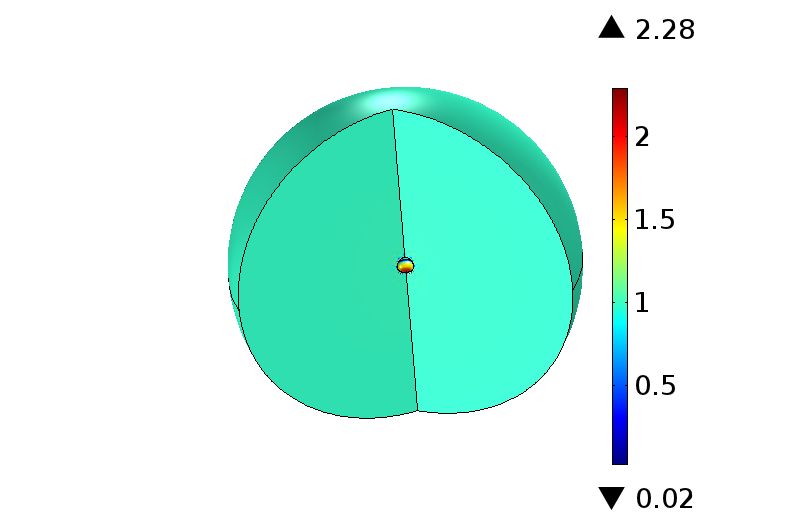}}
    \subfigure{\includegraphics[scale=0.25]{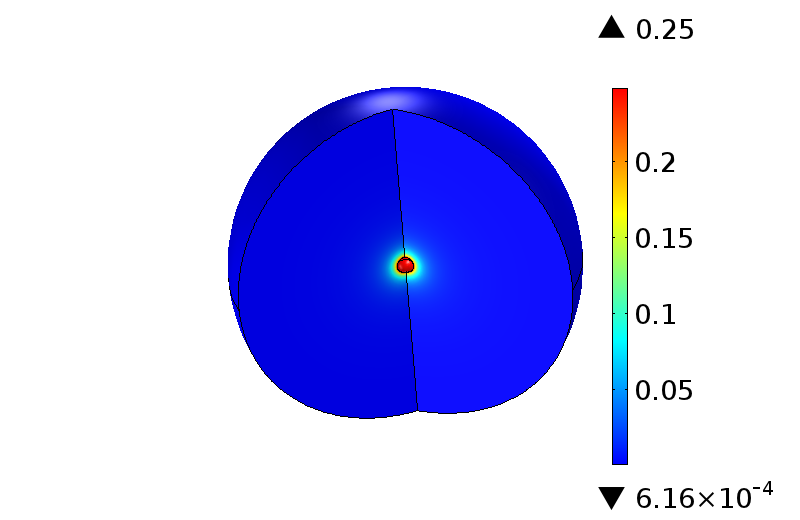}}
    
    {\small (ii) Darcy-Brinkman model.}

	\caption{\textsf{3D axisymmetric problem:} This figure shows the optimal material distribution (left panel), and corresponding pressure (middle) and velocity (right) profiles under the Darcy and Darcy-Brinkman models. Zero body force as well as zero mass source are assumed. The limiting volume of the constrained material (herein, the high-permeability material) is $\gamma = 0.1$ times the total volume of the domain. The regions occupied by the constrained material are shown in `white' while `brown' represents the regions occupied by the unconstrained material (i.e., low-permeability material). (See the online version for the figure in color.) \label{Fig:DB_3D_Axisym_results}}
\end{figure}

Figure \ref{Fig:DB_3D_Axisym_results} shows the material distribution, the pressure and velocity profiles within the 3D spherical domain for the Darcy model and the Darcy-Brinkman model; the parameters used in the numerical simulation are provided in Table \ref{Fig:2D_Circles_BVP_parameters}. The conclusions for this problem are the same as the previous section. The optimal material design under the Darcy-Brinkman model is identical to that of the Darcy model, as even in this problem the viscous shearing stress vanishes due to axisymmetry.


%

\section{OPTIMAL MATERIAL LAYOUTS FOR PIPE-BEND PROBLEM}
\label{Sec:S6_Brinkman_Pipe_bend}
We will use the pipe-bend problem to illustrate the effect of viscous shearing stress on optimal material distribution. This benchmark problem is often used in the computational fluid dynamics literature. For example, \cite{shabouei_nakshatrala_cicp} have used this problem to test the efficacy of their proposed \emph{a posteriori} error measures tailored for flow of fluids through porous media. This benchmark problem has also been used under topology optimization \citep{guest2006topology,borrvall2003topology,phatak2020optimal}; however, the motive of the said works is different from that of ours, which is to compare the optimal material distributions under the Darcy and Darcy-Brinkman models. 

\begin{figure}[h]
  \includegraphics[scale=0.5]{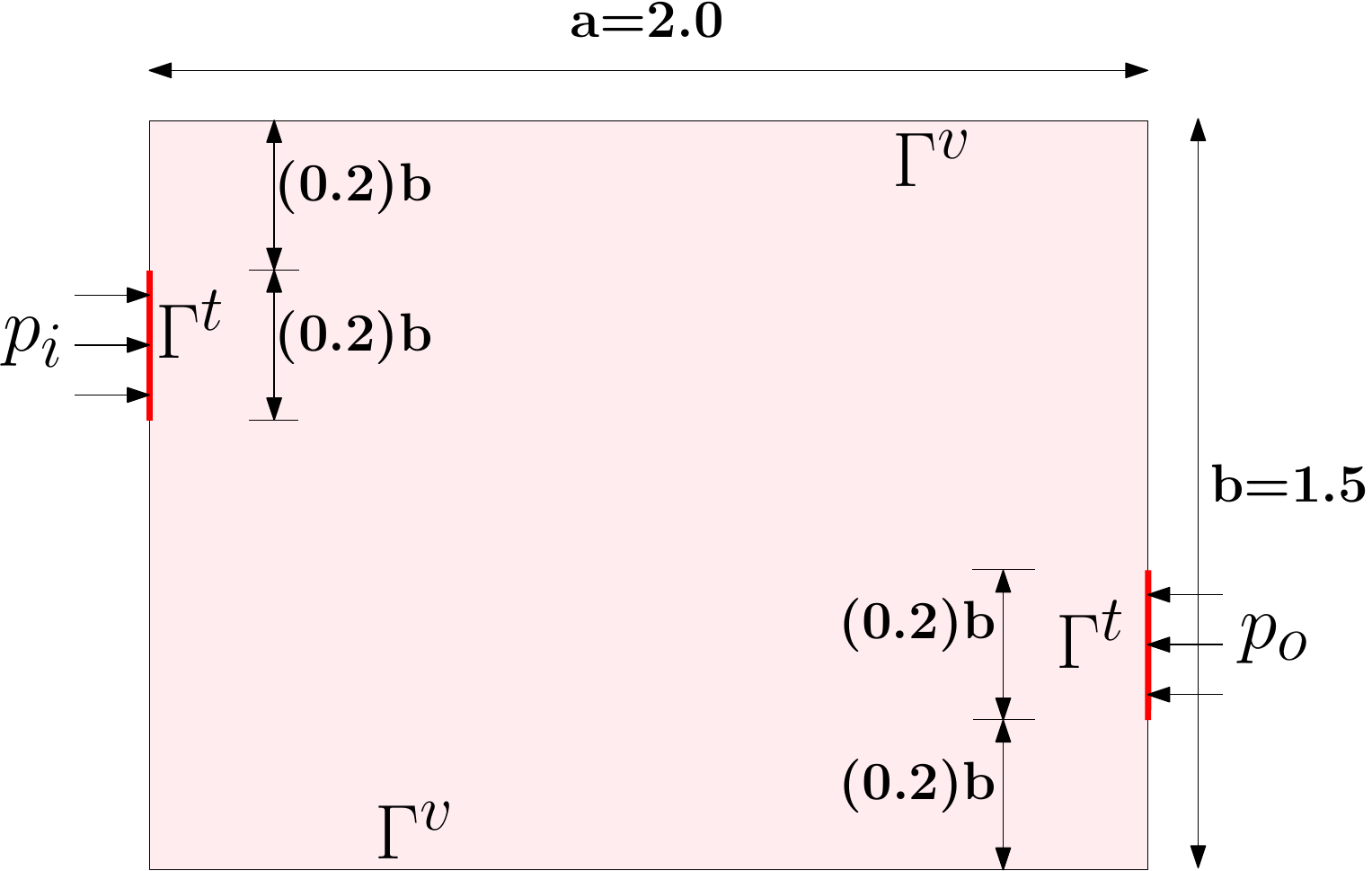}
  \caption{\textsf{2D pipe-bend problem:} This figure shows the geometry and boundary conditions used in the numerical simulation. \label{Fig:PipeBend_BVP}}
\end{figure}

Consider a rectangle domain with dimensions $2 \times 1.5$. The inlet on the left side of the boundary is subjected to a pressure loading of $p_i = 100$. A pressure loading of $p_o = 1$ is applied at the outlet present on the right side of the boundary. (Note the pressure loading will be a traction boundary condition under the Darcy-Brinkman model and a pressure boundary condition under the Darcy model.) The rest of the boundary is subject to homogeneous velocity boundary conditions (i.e., $\mathbf{\mathbf{v}}^\mathrm{p}(\mathbf{x}) = \mathbf{0}$ for the Darcy-Brinkman model, and $\mathrm{\mathbf{v}}(\mathrm{\mathbf{x}}) \cdot \widehat{\mathrm{\mathbf{n}}}(\mathrm{\mathbf{x}}) = 0$ for the Darcy model). Figure \ref{Fig:PipeBend_BVP} provides a pictorial description of the problem for primal analysis.

\begin{figure}[h]
  \subfigure{\includegraphics[scale=0.25]{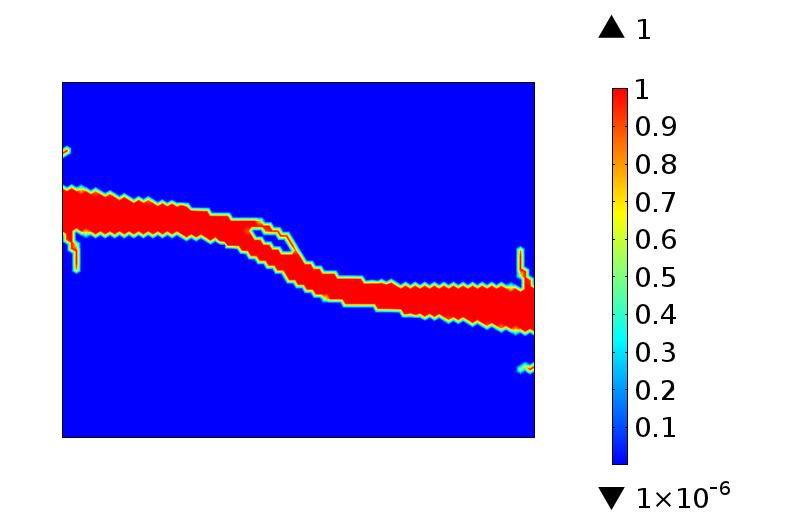}}
  \subfigure{\includegraphics[scale=0.25]{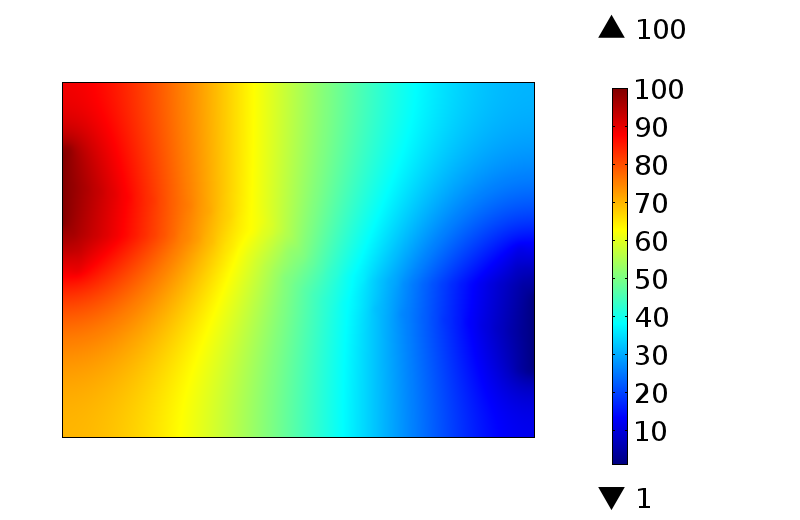}}
  \subfigure{\includegraphics[scale=0.25]{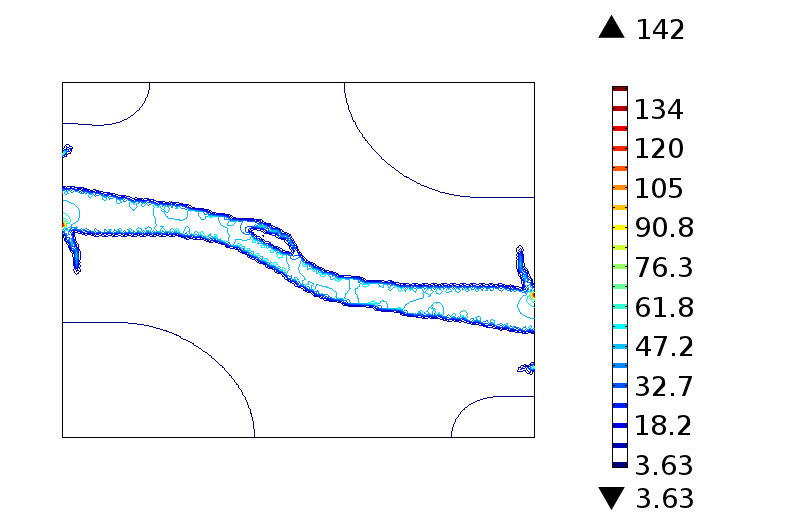}}
  
  {\small (i) Darcy model with $\gamma = 0.1$.}
  
  \subfigure{\includegraphics[scale=0.25]{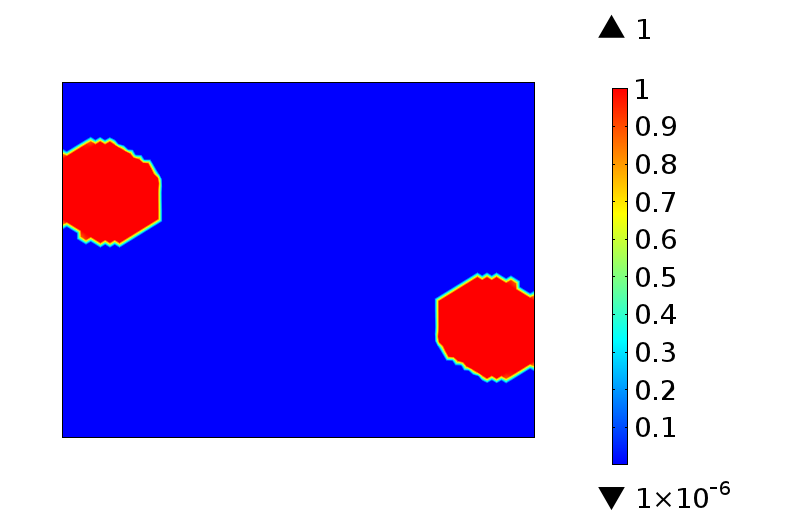}}
  \subfigure{\includegraphics[scale=0.25]{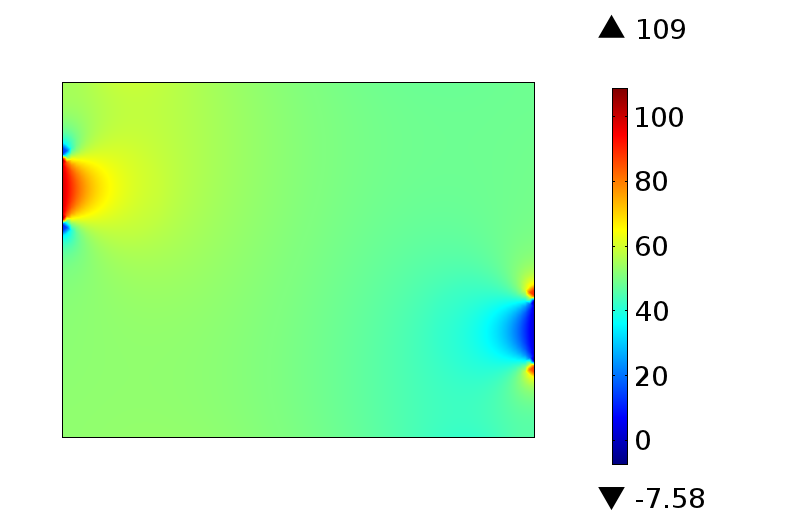}}
  \subfigure{\includegraphics[scale=0.25]{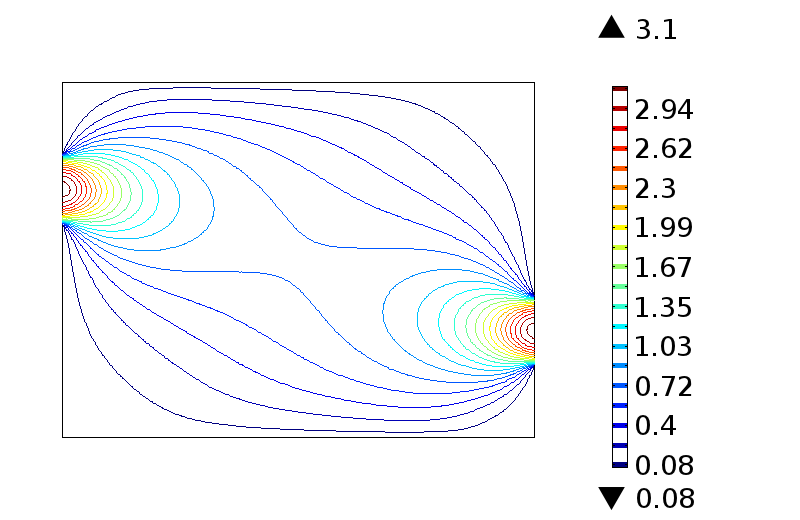}}
  
  {\small (ii) Darcy-Brinkman model with $\gamma = 0.1$.}
  
  \subfigure{\includegraphics[scale=0.25]{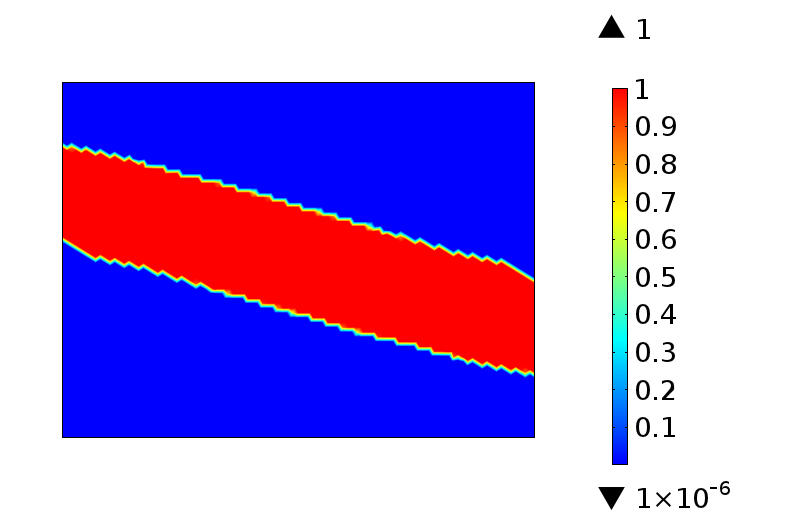}}
  \subfigure{\includegraphics[scale=0.25]{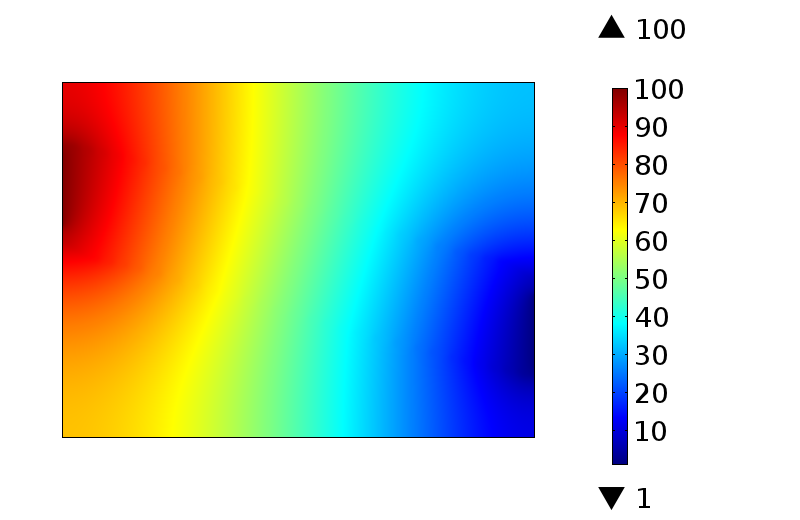}}
  \subfigure{\includegraphics[scale=0.25]{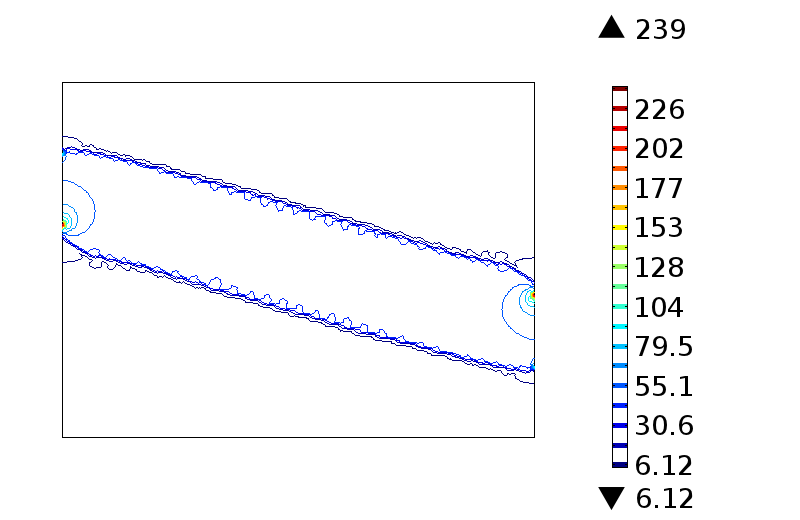}}
  
  {\small (iii) Darcy model with $\gamma = 0.3$.}
  
  \subfigure{\includegraphics[scale=0.25]{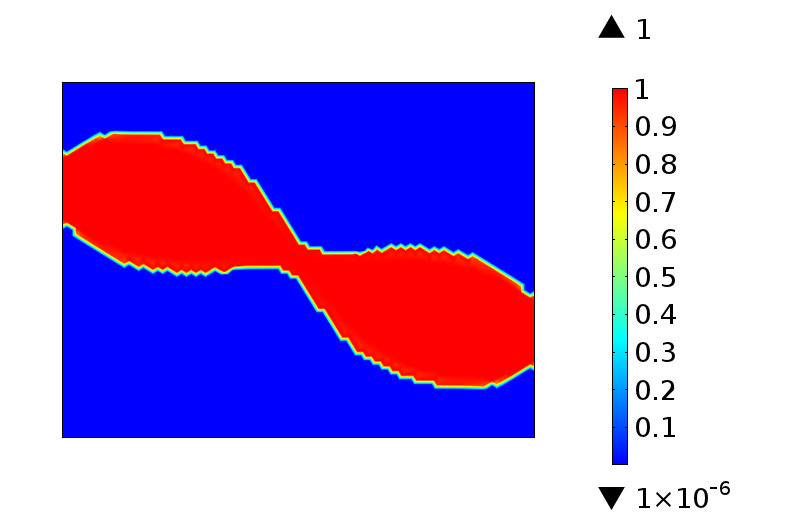}}
  \subfigure{\includegraphics[scale=0.25]{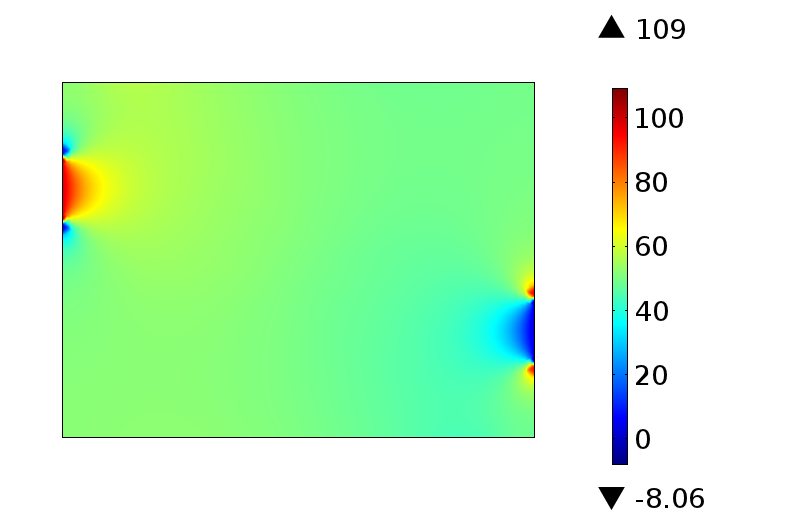}}
  \subfigure{\includegraphics[scale=0.25]{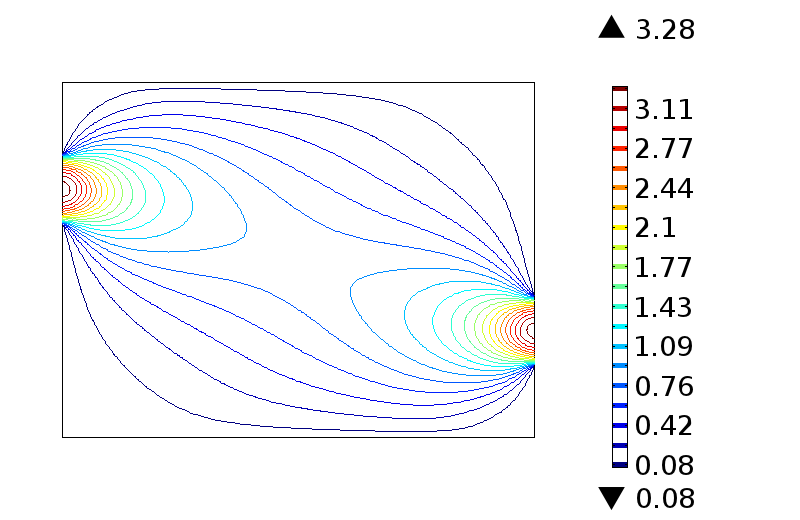}}
  
  {\small (iv) Darcy-Brinkman model with $\gamma = 0.3$.}
	
  \caption{\textsf{Pipe-bend problem:} This figure compares the optimal material distribution (left panel) and the associated pressure (middle) and velocity (right) fields under the Darcy and Darcy-Brinkman models. The results are shown for two values of the volumetric bound constraint: $\gamma = 0.1$ and $0.3$. The computational domain is a rectangle (2.0 x 1.5) with zero body force. The objective is to maximize dissipation with a volume constraint placed on the high permeability material. The constrained (high-permeability) material is represented by `red' while `blue' represents the unconstrained (low-permeability) material. The material distribution as well as the solution fields differ under the two models. (See the online version for the figure in color.) \label{Fig:DB_pipe_bend_results}}
\end{figure}

Table \ref{Fig:2D_Pipebend_BVP_parameters} provides the parameters used in the numerical simulation of the material design problem. Figure \ref{Fig:DB_pipe_bend_results} shows the material design and the associated solution fields for $\gamma = 0.1$ and $0.3$. The main findings from this test problem are: 
\begin{enumerate}
\item Since the flow induces strong viscous shearing stresses, the optimal material distribution under the Darcy-Brinkman model differs from that of the Darcy model. While the Darcy model places the high permeability material along the path connecting the inlet and outlet, the Darcy-Brinkman model places the high permeability material in close proximity to the inlet and outlet.
\item There are marked differences in the solution fields under these models near the inlet and outlet. The solution fields in the interior of the domain, away from the boundaries, are relatively similar under both the models.
\item Even in this problem, the pressure field under the Darcy-Brinkman model does not lie between the pressure loading applied at the inlet $(p_i = 100)$ and outlet $(p_o = 1)$. On the other hand, the pressure field within the domain under the Darcy model lie between the applied pressures at the boundary.
\item An increase in the volume bound constraint $\gamma$ (placed on the high-permeability material) leads to an increase in the flow velocities under both the models; see the velocity profiles in figure \ref{Fig:DB_pipe_bend_results}. This trend makes sense as higher $\gamma$ implies more areal availability for the high-permeability material, thus facilitating a freer passage of the fluid.
\item From Table \ref{Table:DB_pipe_bend_dissipation}, one can notice that the maximum dissipation under the Darcy-Brinkman model is lower than the Darcy model. A complete explanation for this observation will be addressed in a sequel paper.
\end{enumerate}

This problem clearly reveals that the viscous shearing stresses greatly affect the optimal distribution of the porous materials. These stresses, as is the case in this problem, are dominant near the inlet and outlet, especially, when the inlet and outlet are not aligned along the same line.

\begin{table}[h]
  \caption{This table provides the parameters used in the
    numerical simulations of the 2D pipe-bend problem.
    \label{Fig:2D_Pipebend_BVP_parameters}}
  \begin{tabular}{|lr||lr|}\hline
    parameter & value & parameter & value \\ \hline
    $r_i$ & 0.1 & $r_o$ & 1 \\
    $p_i$ & 100 & $p_o$ & 1 \\
    $k_L$ & 0.1 & $k_H$ & 1 \\
    $\gamma$ & 0.1, 0.3 & $\mu$ & 1 \\ \hline
  \end{tabular}
\end{table}

\begin{table}
  \caption{The rate of dissipation for different values of $\gamma$ (i.e., the volume bound constraint) under the Darcy and Darcy-Brinkman models.\label{Table:DB_pipe_bend_dissipation}}
  \begin{tabular}{lcc} \hline 
    Model & $\Phi$ for $\gamma = 0.1$ & $\Phi$ for $\gamma = 0.3$ \\ \hline  
    Darcy model & 1081.1 & 2279.5 \\
    Darcy-Brinkman model & 73.0 & 77.1 \\ \hline 
  \end{tabular}
\end{table}

\section{OPTIMAL MATERIAL DESIGN FOR FLOWS IN A BACKWARD-FACING STEP}
\label{Sec:S7_Backward_Facing_Step}

The backward-facing step is a widely used test problem to study separation of flows due to an abrupt change in geometry \citep{armaly1983experimental,lee1998experimental}. Herein we will use this problem to understand the effect of geometrical changes in tandem with strong viscous shearing stresses on the optimal material distribution.

\begin{figure}[h]
  \includegraphics[scale=0.8]{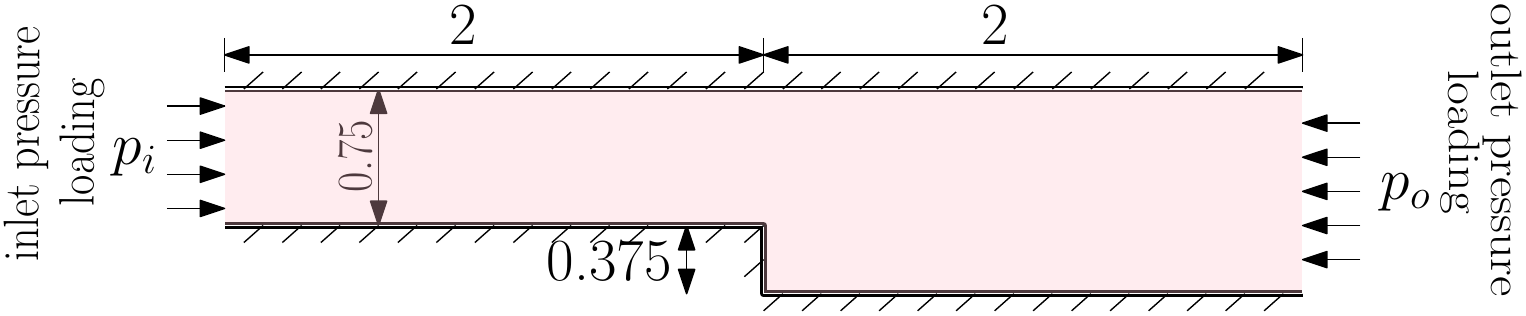}
  \caption{\textsf{Backward-facing step problem:} This figure provides a pictorial description of the boundary value problem, including the dimensions and boundary conditions. The inlet at the left boundary is subject to a pressure loading $p_i$, and the outlet at the right boundary is subject to a pressure loading $p_o < p_i$. Homogeneous velocity boundary conditions are enforced on the rest of the boundary. \label{Fig:DB_Backward_Step_BVP}}
\end{figure}

Consider a two-dimensional domain with an expansion ratio of 1:1.5 as shown in figure \ref{Fig:DB_Backward_Step_BVP}. Fluid enters from the left into a domain geometry of $2 \times 0.75$ followed by a downward step of height $0.375$ with a resulting geometry of $2 \times 1.125$. Fluid exits at the right boundary. The prescribed pressure loadings at the inlet and outlet are $p_i = 100$ and $p_o = 1$, respectively. Homogeneous velocity boundary conditions are enforced on the rest of the boundary. The body force is neglected. In this section, we will get the optimal material design (i.e., solve the material design problem) for flow of fluids through a porous domain with a backward-facing step. 

\begin{figure}[h]
  \subfigure{\includegraphics[scale=0.25]{Figures/Backward_facing_step/Bk_Stp_MaxDiss_Darcy_K1-0_1_Gamma0_1_Rev.png}}
  \subfigure{\includegraphics[scale=0.25]{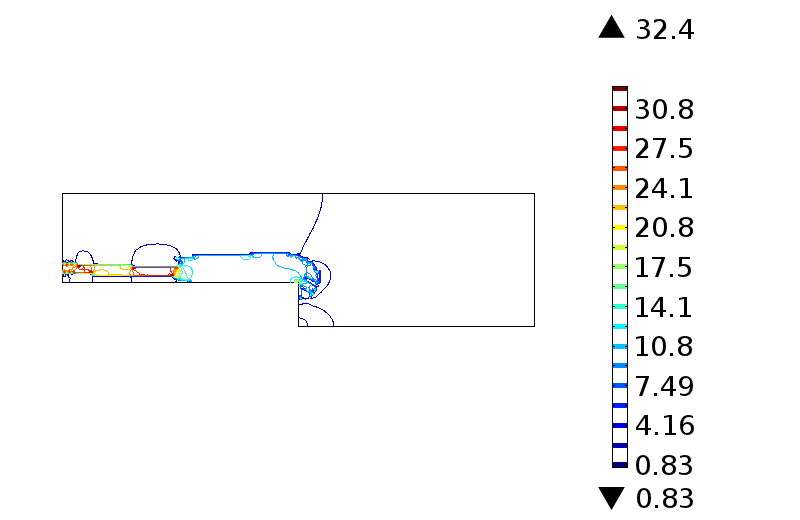}}
  \subfigure{\includegraphics[scale=0.25]{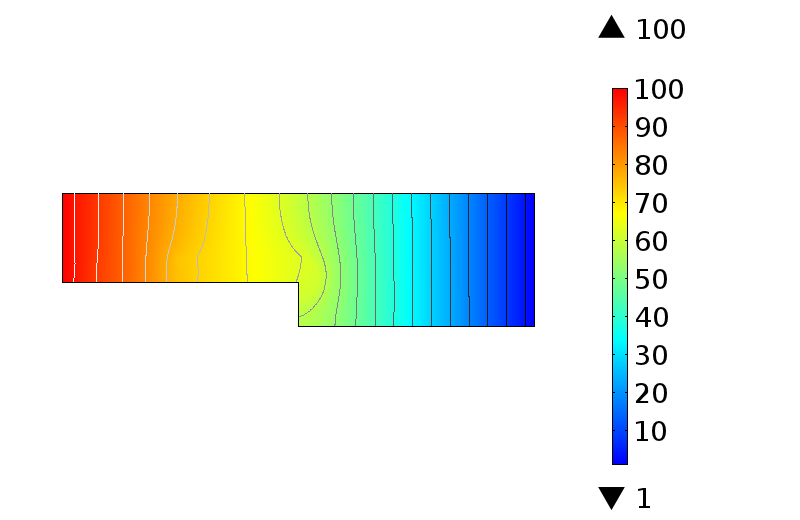}}
  
  {\small \emph{Top panel:} Material distribution (left), velocity (middle) and pressure (right) profiles under the Darcy model.}
  
  \subfigure{\includegraphics[scale=0.25]{Figures/Backward_facing_step/Bk_Stp_MaxDiss_Brinkman_K1-0_1_Gamma0_1_Rev.png}}
  \subfigure{\includegraphics[scale=0.25]{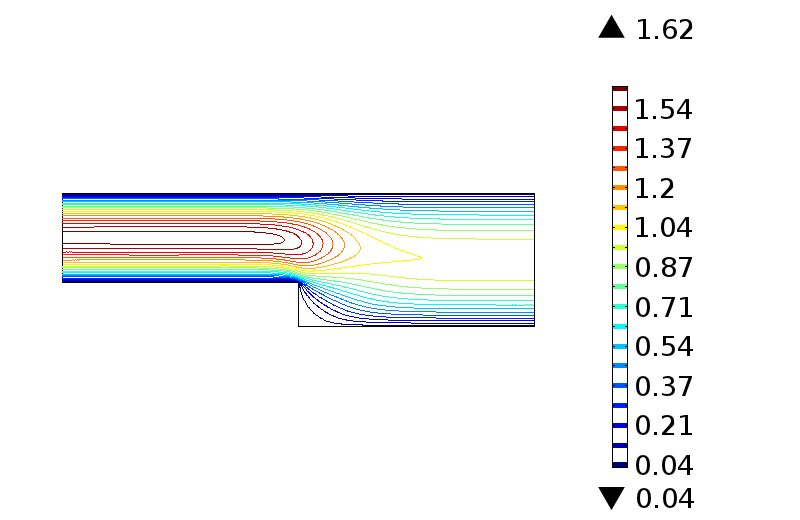}}
  \subfigure{\includegraphics[scale=0.25]{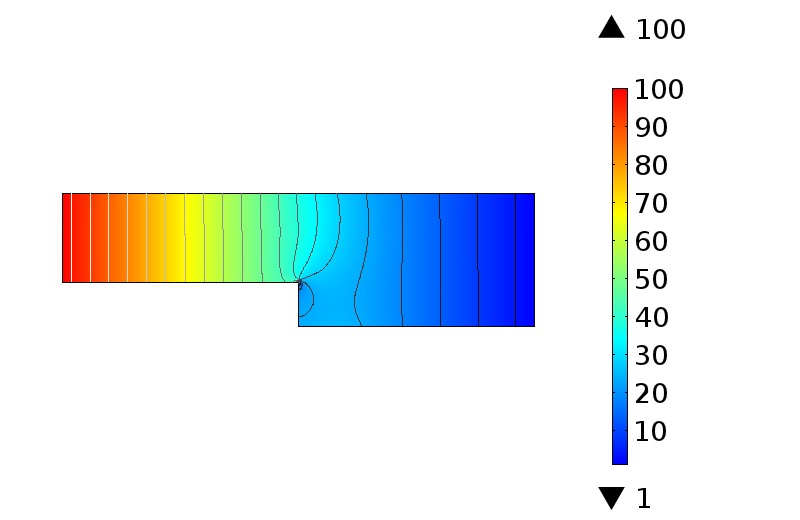}}
  
  {\small \emph{Bottom panel:} Material distribution (left), velocity (middle) and pressure (right) profiles under the Darcy-Brinkman model.}
  
  \caption{\textsf{Backward-facing step problem with $\gamma = 0.1$:} This figure compares the Darcy and Darcy-Brinkman models, contrasting the obtained material designs and solution fields. The limiting area for the constrained (i.e., high-permeability) material is taken as $\gamma = 0.1$. In the material designs, shown in the left panel of the figure, the `red' color denotes the regions occupied by the constrained material while `blue' represents the regions with the unconstrained (i.e., low permeability) material. (See the online version for the figure in color.) \label{Fig:DB_Backward_Step_results_gamma_dot1}}
\end{figure}

\begin{figure}[h]
	\subfigure{\includegraphics[scale=0.25]{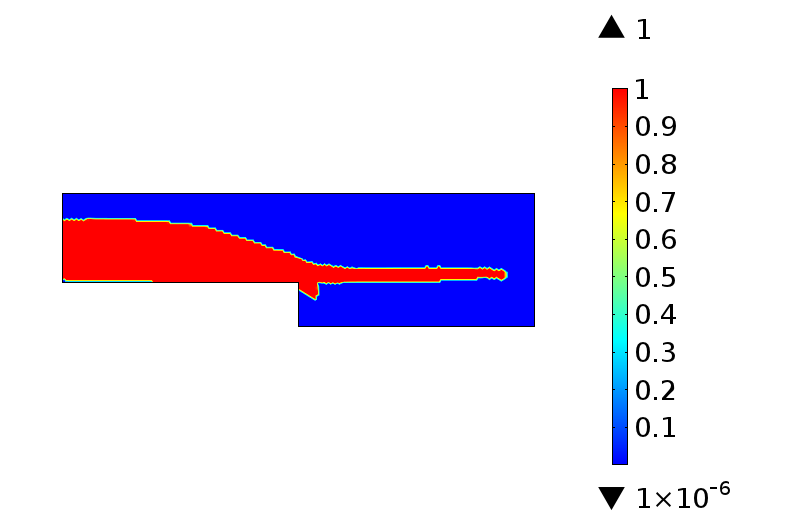}}
	\subfigure{\includegraphics[scale=0.25]{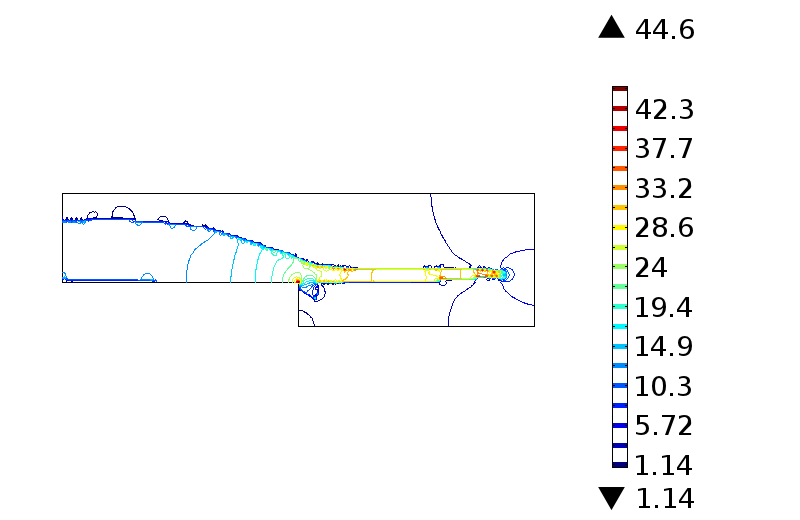}}
	\subfigure{\includegraphics[scale=0.25]{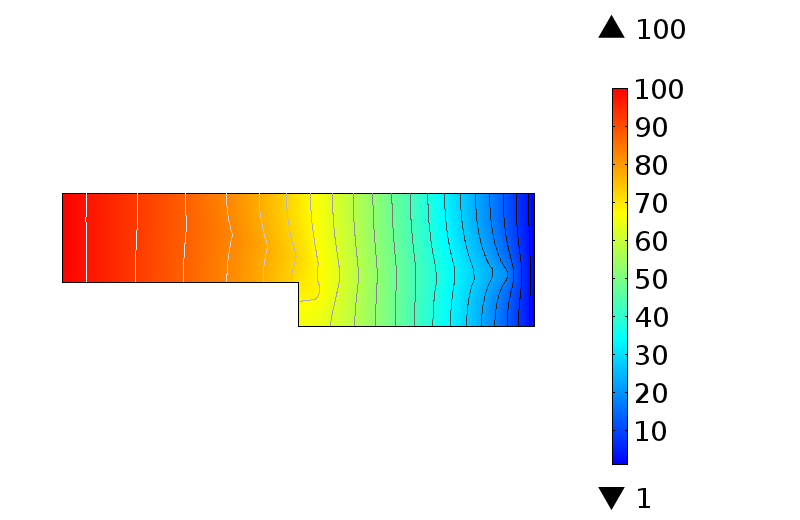}}
	
	{\small \emph{Top panel:} Material distribution (left), velocity (middle) and pressure (right) profiles under the Darcy model.}
	
	\subfigure{\includegraphics[scale=0.25]{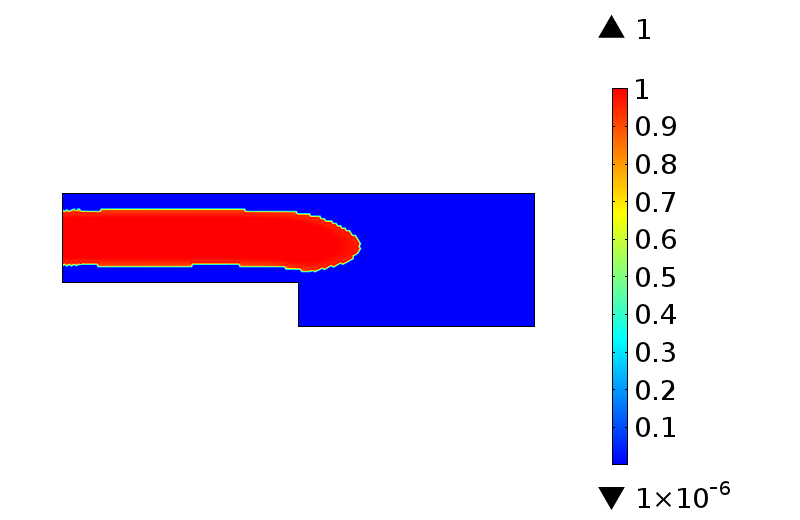}}
	\subfigure{\includegraphics[scale=0.25]{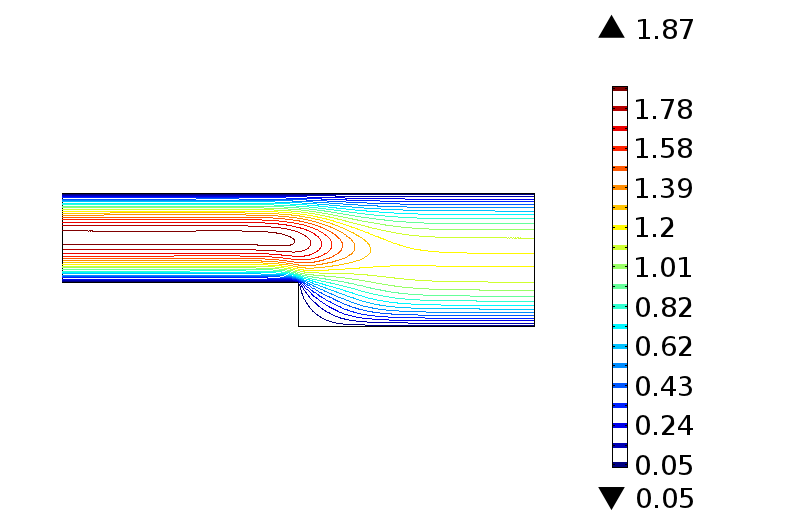}}
	\subfigure{\includegraphics[scale=0.25]{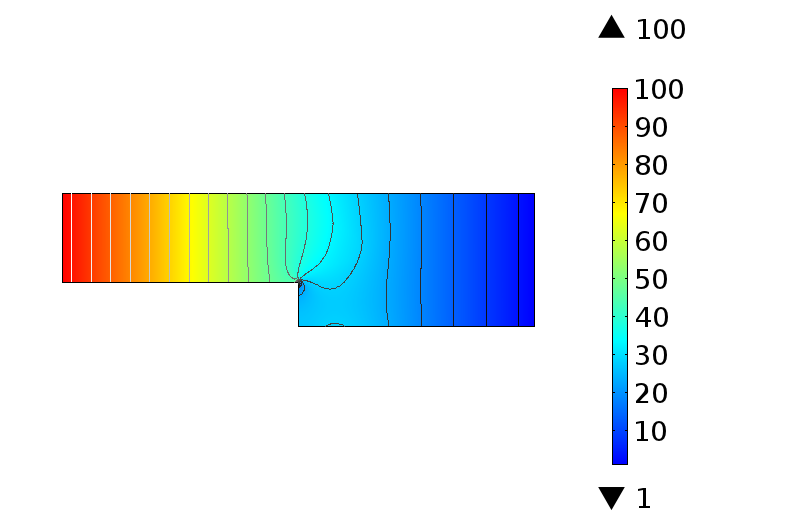}}
	
	{\small \emph{Bottom panel:} Material distribution (left), velocity (middle) and pressure (right) profiles under the Darcy-Brinkman model.}
	
	\caption{\textsf{Backward-facing step problem with $\gamma = 0.3$:} This figure compares the Darcy and Darcy-Brinkman models, contrasting the obtained material designs and solution fields. The limiting area for the constrained (i.e., high-permeability) material is taken as $\gamma = 0.3$. In the material designs, shown in the left panel of the figure, the `red' color denotes the regions occupied by the constrained material while `blue' represents the regions with the unconstrained (i.e., low permeability) material. (See the online version for the figure in color.) \label{Fig:DB_Backward_Step_results_gamma_dot3}}
\end{figure}

This problem is not amenable to an analytical solution for the either primal analysis or material design. We, therefore, resorted to numerics. Table \ref{Fig:Backward_step_BVP_parameters} provides the parameters used in the numerical simulation. (The high and low permeabilities are denoted by $k_H$ and $k_L$, respectively.) Figures \ref{Fig:DB_Backward_Step_results_gamma_dot1} and \ref{Fig:DB_Backward_Step_results_gamma_dot3} show the optimal material distribution and the associated solution fields under both the models for two different values of volumetric bound constraint: $\gamma = 0.1$ and $0.3$.

\begin{table}[h]
  \caption{\textsf{Backward-facing step problem:}~This table provides the parameters used in the numerical simulation.
    \label{Fig:Backward_step_BVP_parameters}}
  \begin{tabular}{|lr||lr|}\hline
    parameter & value & parameter & value \\ \hline
    $r_i$ & 0.1 & $r_o$ & 1 \\
    $p_i$ & 100 & $p_o$ & 1 \\
    $k_L$ & 0.1 & $k_H$ & 1 \\
    $\gamma$ & 0.1 & $\mu$ & 1 \\ \hline
  \end{tabular}
\end{table}

The numerical results in these figures illustrate four salient features: (i) the material designs under the two models differ significantly, as the abrupt change in the geometry results in significant viscous shearing stresses.  (ii) Due to the difference in the material distributions, the associated velocity fields differ qualitatively and quantitatively (i.e., an order of magnitude) for these two models. (iii) The pressure within the domain under the Darcy-Brinkman model can be lower than the prescribed pressure loading at the inlet. On the other hand, the pressure within the domain under the Darcy model lies between the prescribed pressure loadings on the boundary (i.e., $p_o \leq p(\mathbf{x}) \leq p_i$). (iv) As seen even in the previous problem, the magnitude of the velocity increases as $\gamma$ (which is limits the area occupied by the high-permeability material) increases. 


\section{CLOSURE}
\label{Sec:S8_Brinkman_CR}
This paper studied optimal material layouts under the Darcy-Brinkman
model and compared these layouts with the ones obtained under the Darcy
model. Topology optimization is used to get these optimal material designs.
The rate of dissipation---a physical quantity with firm thermodynamic
basis---is used to define the objective function. Since we considered
pressure-driven problems, the rate of dissipation is maximized with a
volumetric bound constraint on the high-permeability material. We obtained
analytical solutions for 2D and 3D axisymmetric problems; these solutions
will be valuable to verify numerical simulators for topology optimization.

Based on the analytical and numerical solutions for the design problem under various boundary value problems, presented in the previous sections, answers to the questions laid in Introduction (\S\ref{Sec:DB_Intro}) are as follows:
\begin{enumerate}[(C1)]
\item In general, the optimal material layouts under the Darcy-Brinkman model differ from that of the Darcy model. Moreover, due to different material designs, the associated solution fields (i.e., pressure and velocity) are qualitatively and quantitatively different for the two models.
\item The said difference in material layouts is prominent for those problems that exhibit high viscous shearing stresses. Thus, viscous shearing stresses significantly affect the material design for applications involving the flow of fluids through porous media.
\item Flows through domains with abrupt changes in the geometry (e.g., a reentrant corner) or the presence of close-by boundaries will experience strong viscous shearing forces. So, the domain's geometry will affect the material design and the choice of the model for the primal analysis. 
\item For the class of problems exhibiting axisymmetry, the material distributions under the Darcy-Brinkman and Darcy models are identical. The reason is that viscous shearing stress vanishes for these problems. However, the solution (i.e., pressure and velocity) fields within the domain will not be the same.
  \item To address the question---which model to use for a given problem---we offer the following guidelines:
  \begin{enumerate}[(i)]
  \item Since the Darcy-Brinkman model considers the internal friction within the fluid, the model is capable of accurately capturing viscous shearing stresses, which occur near solid surfaces. Thus, for the problems with close-by boundaries and domains with blunt objects, use the Darcy-Brinkman model.
 \item However, problems exhibiting axisymmetry, for which viscous shearing stresses vanish, the optimal material layouts under these two models are identical. So, for axisymmetric problems, one can use the Darcy model, which is the simpler of the two models.
\item In the absence of such apparent features, our suggestion is to perform a primal analysis (using one of the given porous materials in the entire domain) on the boundary value problem and determine whether the viscous shearing stresses are dominant. If so, use the Darcy-Brinkman model. Otherwise, use the Darcy model, which is easier to solve numerically because of its simplicity.
\end{enumerate}
\end{enumerate}
  
Other important observations are:
\begin{enumerate}[(O1)]
\item Under the Darcy model, the pressure field within the domain near boundary matches the prescribed pressure loadings. However, this trend may not be true under the Darcy-Brinkman model. The reason is that the definition for the traction under the Darcy-Brinkman model involves the gradient of the velocity field besides the pressure field. 
\item Increasing the value of $\gamma$, the volumetric bound constraint that limits the amount of area/volume occupied by the high-permeability, increases the magnitude of the velocity.
\item We have observed that, for the problems considered in this paper, the rate of dissipation for the optimal material distribution under the Darcy-Brinkman model is lower than that of the Darcy model. This trend has a deeper reason, valid even for the primal analysis and not just for the design problem; hence, a separate in-depth study to explain the mentioned trend.
\end{enumerate}

  In conclusion, the selection of a model for primal analysis impacts the (material) design problem; hence, its selection should be done with care. A plausible future work can be towards providing a scientific explanation for the trend reported in (O3) about the relative magnitudes of the rate of dissipation under the two models for a given boundary value problem.

\bibliographystyle{plainnat}
\bibliography{Master_References}


\end{document}